\documentclass[aps,twocolumn,prb,showpacs]{revtex4}

\usepackage{graphicx}
\usepackage{amsmath}

\newcommand{\be}{\begin{equation}}
\newcommand{\ee}{\end{equation}}

\begin{document}

\title{Frenkel-Kontorova Models, Pinned Particle Configurations and Burgers Shocks}

\author{Muhittin Mungan$^{1,2}$\footnote{ {\tt mmungan@boun.edu.tr}} and Cem Yolcu$^3$\footnote{ {\tt yolcu@cmu.edu}}} 
\affiliation{$^1$Department of Physics, Faculty of Arts and Sciences,\\
Bo\u gazi\c ci University, 34342 Bebek, Istanbul, Turkey,}
\affiliation{$^2$The Feza G\"ursey Institute,
P.O.B. 6, \c Cengelk\"oy, 34680 Istanbul, Turkey,}
\affiliation{$^3$ Department of Physics, Carnegie Mellon University, Pittsburgh PA 15213, USA}

\date{\today}

\begin{abstract}

We analyze the relationship between the lowest energy configurations of 
an infinite harmonic chain of particles in a periodic potential and the evolution of 
characteristics in a periodically-forced inviscid Burgers equation. 
The shock discontinuities in the the Burgers evolution arise from thermodynamical 
considerations and play an important role as they separate out flows related to 
lowest energy configurations from those associated with higher energies. We study in detail 
the exactly solvable case of an external potential consisting of parabolic segments, 
and calculate analytically the lowest energy configurations, as well as excited states 
containing  discommensurations. 

\end{abstract}



\maketitle

\section{Introduction}

The Frenkel-Kontorova (FK) model is a classical infinite chain of atoms linked by elastic springs with equilibrium spacing $\mu$, subject to an external periodic potential of period $2a$  \cite{FK}. It is a simple model for the description of dislocations in solids but has also been applied to interfacial slip in  
earthquakes\cite{WeissElmer,CuleHwa,Gershenzon}, the dynamics of DNA denaturation \cite{PB}, as well as other areas \cite{Braun}.
The FK model is characterized by the competition of two different length scales, $\mu$ and $2a$, and 
two energy scales set by the external potential and the spring elastic energy.  
The static configurations of lowest energy have a rather complex dependence on these parameters:  
they can be commensurate or incommensurate with the period of the external potential \cite{Aubry} and 
the transition between them is a critical phenomenon exhibiting scaling \cite{SheKad,Sue,PeyAub,MacKay}. The FK model has been analyzed based on a connection with two dimensional area-preserving maps, 
such as the standard map \cite{Aubry}. Aubry\cite{Aubry1983} and Mather\cite{Mather} have shown that the lowest energy configurations correspond to invariant sets of the associated maps that can be KAM tori, Cantor sets (Cantori), or limit cycles corresponding respectively to unpinned incommensurate, pinned incommensurate and pinned commensurate configurations. 

It was recently discovered that a continuum hydrodynamic type of evolution underlies 
the FK models \cite{JKM,EKMS}. For the case of an elastic chain of particles 
embedded in an external potential, this evolution is governed by a 
periodically-forced inviscid Burgers equation and the associated flow of characteristics turns out to be closely related to the particle configurations. This connection was further developed by E and Sobolevskii \cite{E,Sobolevskii} (see also the review by Bec and Khanin \cite{BecKhanin}).

The purpose of the present article is to illustrate and further investigate the relation between FK models and its description in terms of a forced Burgers equation by explicitly working out an example. 
This is desirable for two reasons: On the one hand, the results obtained in \cite{JKM,EKMS,E,Sobolevskii} are mathematical, centering mostly around existence theorems and properties of the flow with less emphasis on connections with particular physical models. It would therefore be useful to consider an example
that can be exactly solved using this approach, and thereby 
illustrate explicitly how the flow properties relate to physical properties of the FK model such 
as the lowest energy configurations and excited states. On the other hand, 
most of the results on FK models focus on the lowest energy configuration from which other properties such as the stability region of a given configuration can 
be obtained. However, 
the Burgers description and in particular the flow patterns provide additional insight, allowing us to construct analytically the excited states with and without discommensurations and to see how the flow pattern changes as the parameters of the model are varied, particularly near the boundary of a region of stability. 

The description of FK models by an underlying one parameter continuum flow also constitutes a novel technique which should be applicable to the larger class of problems \cite{WeissElmer,CuleHwa,Gershenzon} mentioned above, as well 
as FK models with more complex external potentials \cite{GSU,LT}. 
The article is organized as follows. In Section \ref{FKReview} we briefly review the basic results due to Aubry, Aubry and collaborators, and Mather \cite{Aubry, Aubry1983, Aubry1983a,Mather}. We then proceed in Section \ref{BurgersIntro} to give a simple derivation of the relation between FK models and the periodically forced Burgers equation and discuss general properties of the solutions. In Section \ref{FKwithScalloped}, we focus on the FK model with a piece-wise parabolic potential. This model is interesting in its own right, as it is related to other FK models in the limit of strong external potential where the particles are confined to regions close to the potential minima. 
We conclude with a discussion of our 
results and possible generalizations. Necessary background material as well as details of some 
of the calculations have been gathered in the appendix. 
 
\section{Infinite Harmonic Chains in Periodic Potentials}
\label{FKReview}

We consider the static lowest energy configurations of an infinite chain of particles of unit mass 
connected by Hookean springs with equilibrium spacing $\mu$ and subject to a periodic external 
potential with period $2a$. The energy of a configuration $\{ y_i\}$ is given by the Hamiltonian 
\begin{equation}
 \mathcal{H}(\{y_i\}) = \sum_{i=-\infty}^{\infty} \left[\frac{1}{2\tau} \left (y_{i+1} - y_i - \mu \right )^2 + V(y_i)\right], 
\label{eqn:Hx}
\end{equation}
where for reasons that will become apparent soon, we have written the spring-constant as $1/\tau$.

Any equilibrium configuration must satisfy the set of coupled difference equations
\begin{equation}
y_{i+1} - 2y_i + y_{i-1}  - \tau V^\prime(y_i) = 0,
\label{eqn:DE}
\end{equation}
but solutions to the above equation will in general not be lowest energy configurations. 

Notice that due to the periodicity of the external potential $V$, if $\{ y_i\}$ is an equilibrium 
configuration, so is the configuration $\{ y_i + 2a \}$, where each particle has been shifted by an 
amount of $2a$. Making the change of variables $\tilde{y}_i = y_i \; {\rm mod} \; 2a$ and 
$p_{i+1} = \tilde{y}_{i+1} - \tilde{y}_{i}$, Eq.~(\ref{eqn:DE}) becomes
\begin{eqnarray}
\tilde{y}_{i+1} &=& \tilde{y}_i + p_{i+1}, \label{eqn:HM1} \\
p_{i+1} &=& p_i + \tau V^\prime(\tilde{y}_{i}), \label{eqn:HM2}
\end{eqnarray}
which is a 2d Hamiltonian mapping on the cylinder $\mathcal{S} \times \mathcal{R}$  
called a {\em twist map}.  In the case of a sinusoidal external potential $V(y) = 1 - \cos (\pi y /a)$, this map is also known as the  {\it standard map} \cite{Aubry,JoseSaletan}.

Aubry and Mather have independently shown that the lowest energy configurations correspond to special invariant sets  of the twist map, Eqs.~(\ref{eqn:HM1}-\ref{eqn:HM2}), with limit cycles corresponding to commensurate structures, whereas trajectories that are dense on KAM tori correspond to incommensurate structures \cite{Aubry}. The KAM theorem applied to twist maps  indicates that there is a 
critical threshold $\tau_c$ such that for $\tau > \tau_c$ all KAM tori have broken up and the incommensurate 
structures are dense on Cantor sets (Cantori), that have measure zero. Thus for $\tau > \tau_c$ almost all 
lowest energy structures are commensurate (strong pinning limit). 

Aubry has shown that under general conditions \cite{AubryLeDaeron} to each lowest energy configuration $\{y_i\}$ there is associated a winding number $\ell$ given by 
\begin{equation}
\ell  = \lim_{N - N^\prime \rightarrow \infty} \frac{y_{i+N} - y_{i+N^\prime}}{N - N^\prime}, 
\label{eqn:ell}
\end{equation}
which is the average distance between two neighboring particles in the lowest energy configuration. 
Moreover, $\ell$ as a function of the 
parameters $\mu$ and $\tau$ takes constant values for each rational value of 
$\ell/2a$. Hence this function is a {\it Devil's staircase}. Aubry has also shown that the lowest energy 
configurations $\{y_i\}$ are of the form
\begin{equation}
y_i = f(i \ell + \alpha) = i \ell + \alpha + g(i \ell + \alpha), 
\label{eqn:fgeq}
\end{equation}
where $g$ is periodic with period $2a$ and the choice of $\alpha$ only serves to determine which particle on the infinite chain is to be denoted the zeroth particle. The function $g$, or equivalently $f$, is 
called the hull function \cite{Aubry}. In the presence of KAM tori the hull function is continuous,  whereas for a given $\ell$ and sufficiently large $\tau$ it becomes discontinuous.

\section{FK Models and Characteristic Flows in a Periodically-Forced Inviscid Burgers Equation}
\label{BurgersIntro}

The determination of the lowest energy configurations as sketched in the previous section is rather indirect. The equilibrium equations Eq.~(\ref{eqn:DE}), or Eqs.~(\ref{eqn:HM1})-(\ref{eqn:HM2}), do not directly yield them. Rather, the lowest energy configurations turn out to correspond to a highly special subset of orbits of the mapping. Their properties can be described qualitatively from the general properties of twist maps. Analytical calculations for a particular model are hard, and numerical simulations are difficult, since the trajectories in question are hyperbolic and hence numerically unstable \cite{Greene,Sue}. 
It would therefore be desirable to obtain such configurations in a more direct way analytically.

\subsection{The Fundamental Catastrophe}
\label{ssec:Burgers}

\begin{figure}
\includegraphics[height=8cm]{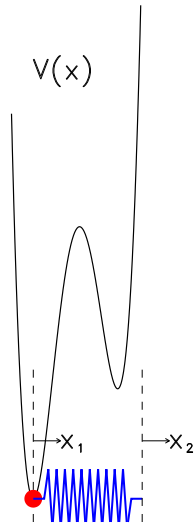}
\caption{Particle in an external potential $V(x)$ connected to a spring with stiffness 
$1/\tau$. The positions of the particle and the endpoint of the spring are given by $x_1$ and 
$x_2$, respectively, such that $x_1 = x_2$ corresponds to the spring being unstretched. }
\label{fig:cat1}
\end{figure}

Consider a particle in an arbitrary potential (not necessarily periodic), as shown in 
Fig.~\ref{fig:cat1}, to which there   
is connected a spring of spring constant $1/\tau$. Let $x_1$ denote the location of the particle, 
while $x_2$ denotes the position of the endpoint of the spring, such that $x_1 = x_2$ corresponds 
to the spring being unstretched. 

\begin{figure*}
\includegraphics[height=5.7cm]{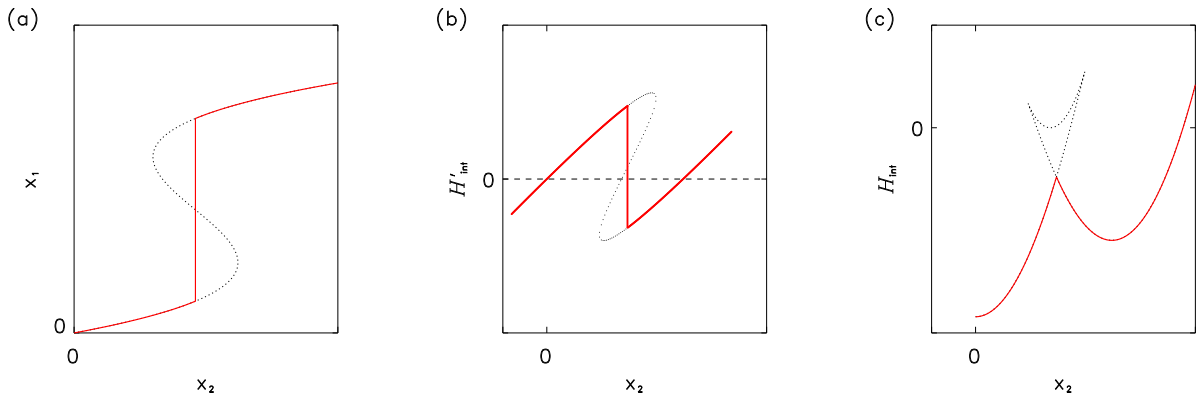}
\caption{(Color online) (a) The Lagrange map yielding the position $x_1$ of the particle in Fig.~\ref{fig:cat1} as a function of the endpoint of the spring $x_2$. At a certain $x_2$ the particle abruptly jumps wells. The dotted line is a plot of Eq.~(\ref{eqn:x2vsx1}). The actual location of the discontinuity on the red curve is obtained from an adiabatic condition. (b) The derivative of the internal energy of the particle $H_{\rm int}$ as a function of the endpoint $x_2$ of the string. The dotted curve is the multiple-valued solution 
$H^\prime_{\rm int}$ containing overhangs. The location of the discontinuity  
$x_2$ is such that the areas bounded by the dotted curves and 
are equal. (c) The internal energy $H_{\rm int}$ along the corresponding two paths in (a). }
\label{fig:cat2}
\end{figure*}

We are interested in the position of the particle $x_1$ as a function of 
$x_2$. This can be found from minimizing the Hamiltonian
\begin{equation}
H(x_2,x_1) = \frac{1}{2\tau} (x_2 - x_1)^2 + V(x_1) 
\label{eqn:cat1}
\end{equation}
with respect to $x_1$ so that 
\begin{equation}
x_1 =  \arg \min_{x_1}  H(x_2,x_1) 
\label{eqn:cat2}
\end{equation}
The problem is non-trivial due to the non-convexity of $V(x)$. Differentiating $H(x_2,x_1)$ with 
respect to $x_1$, 
\begin{equation}
 x_2 = x_1 + \tau V^\prime(x_1). 
\label{eqn:x2vsx1}
\end{equation}
For $\tau$ sufficiently large this equation does not necessarily have a unique solution for all values of $x_2$ anymore, 
as the dotted curve in Fig.~\ref{fig:cat2}(a) shows. In order to obtain a single-valued 
dependence of $x_1$ on $x_2$ an additional assumption is needed:
we require that the work done in moving the end point of the 
spring is equal to the change in total internal energy of the particle. The latter is given by 
\begin{equation}
H_{\rm int}(x_2,\tau) \equiv H(x_2,x_1 )\vert_{x_1 = x_1(x_2,\tau)}
\label{eqn:Hintx}
\end{equation}
resulting in the red curve in Fig.~\ref{fig:cat2}(a) with a jump discontinuity which  
corresponds to the particle abruptly switching wells as $x_2$ is increased. 
As a result, the jump of the particle from one well to the other does not 
generate any heat and the process is adiabatic.  The assumption made is thus thermodynamical.  
The work done on the system is readily shown to be continuous in $x_2$, and    
$H_{\rm int}(x_2,\tau)$ is continuous in $x_2$ as well.  In terms of the location of the jump 
discontinuity this implies that the areas bounded by the dashed curve and the discontinuity in Fig.~\ref{fig:cat2}(b), have to be equal. This is the familiar Maxwell 
equal-area construction. 

Let us now reconsider the internal energy of the system by treating the reciprocal spring constant $\tau$ 
as an additional variable. By definition,  
$H_{\rm int}(x_2, \tau)$ is a potential so that the associated force  
\begin{equation}
F(x_2,\tau) = - \frac{\partial H_{\rm int}}{\partial x_2}, 
\end{equation}
corresponding to the restoring force at the end point of the spring, must be conservative. 
From simple mechanical considerations it is also clear that 
\begin{equation}
F(x_2,\tau) = - \frac{{\rm d}V}{{\rm d} x_1}
\end{equation}
where $x_1$ satisfies Eq.~(\ref{eqn:x2vsx1}). 
Note in particular, that for $\tau = 0$, corresponding 
to an infinitely stiff spring, we have $x_1 = x_2$, so that 
\begin{equation}
- \frac{{\rm d}V}{{\rm d} x_1} = F(x_1,0).
\end{equation}

Combining the above equations, we find,
\begin{equation}
F(x_2,\tau) = F(x_1,0), \;\; \mbox{where} \;\;\;\ x_2 = x_1 - \tau F(x_1,0).
\label{eqn:FBurgers}
\end{equation}
In other words, $F(x,t)$ remains constant on the line $x = x_1 - t F(x_1,0)$ with $0 \le t < \tau$.
Physically, this is just a restatement of the fact that the forces on both ends of the 
spring are equal. Mathematically, however Eq.~(\ref{eqn:FBurgers}) implies that $F(x,t)$ is 
a solution of the inviscid Burgers equation
\begin{equation}
\frac{\partial F}{\partial t} - F\; \frac{\partial F}{\partial x} = 0
\end{equation}
with initial condition $F(x,0) = -{\rm d} V/{\rm d}x$. 

It is more convenient to write the inviscid Burgers equation in its more familiar form by 
letting $u(x,t) = - F(x,t)$. Denoting partial derivatives by subscripts, 
we have
\begin{equation}
 u_t + u u_x = 0 
\label{eqn:burgers}
\end{equation}
with $u(x,0) = {\rm d} V/{\rm d}x$ and from Eq.~(\ref{eqn:FBurgers}),  
\begin{equation}
 u(x,\tau) = u(x_0,0),
\label{eqn:burgersIC}
\end{equation}
for $x, x_0$ and $\tau$ satisfying the characteristic equation
\begin{equation}
x = x_0 + \tau u(x_0,0),
\label{eqn:burgers_char} 
\end{equation}
which are lines on the $x\tau$ plane. 

Although $u$ is constant on the characteristics, whenever ${\rm d} u/{\rm d}x < 0$, 
characteristic lines intersect, corresponding to multiple-valued solutions. As we have shown, this situation is resolved by the equal-area construction and gives rise to a discontinuity, 
a {\em shock}. Such solutions are called weak since, Eq.~(\ref{eqn:burgers})
can only be satisfied in a weak sense, due to the discontinuities of $u$. 
Further relevant details on weak 
solutions of the inviscid Burgers are provided in Appendix \ref{onBurgers}.  

It is important to stress that the weak solutions are not a mathematical 
artifact, but follow from thermodynamical considerations. To see this more clearly, consider the 
particle-spring system embedded in a heat bath of temperature $T$. The position of the particle 
$x_1$ can be thought of as an internal variable and we consider the partition function
\begin{equation}
e^{-\beta \epsilon(x,\tau)} = \left (\frac{\beta}{2\pi\tau} \right )^{1/2}\; \int 
e^{ -\frac{\beta}{2\tau} (x - x^\prime)^2}\; e^{ - \beta V(x^\prime)}
\; {\rm d}x^\prime
\end{equation}
as a function of the external variable $x$, with $\epsilon(x,\tau)$ being the corresponding 
free energy. An $x$-independent pre-factor can be arbitrarily chosen, but with the 
choice made above $e^{-\beta \epsilon(x,t)}$ is a 
solution of the diffusion equation with diffusion constant $kT/2$. The 
Cole-Hopf transformation\cite{Cole,Hopf} $u(x,t) = \epsilon_x(x,t)$ yields the {\it viscid} 
Burgers Equation for $u$:
\begin{equation}
u_t + uu_x = \frac{kT}{2} u_{xx} 
\label{eqn:viscidBurgers}
\end{equation}
with initial condition $u(x,0) = V^\prime(x)$. The same weak solution also follow 
from the solution of Eq.~(\ref{eqn:viscidBurgers}) at non-zero $T$ in the limit $T \rightarrow 0$ 
\cite{Leveque,Whitham,Evans}. Thermodynamically, this corresponds to cooling the system quasi-statically 
to zero temperature resulting in a lowest energy configuration, {\it cf.} Fig.~\ref{fig:cat2}(c). 
The internal energy $H_{\rm int}(x,\tau)$ is thus the zero-temperature limit of the free energy 
$\epsilon(x,\tau)$.

\subsection{Burgers Evolution of the FK Model}

The construction presented in the previous section can be utilized to treat the lowest energy configurations 
of the FK model with Hamiltonian Eq.~(\ref{eqn:Hx}). Consider a semi-infinite chain with particle configurations $\{y_i\}$ such that  
$ - \infty < i < n$. Denote by $H_{\rm int}^{(n)}(y_n,\tau)$ the energy of a lowest energy configuration 
of the semi-infinite chain with its end point fixed at $y_n$. Owing to the periodicity of the 
external potential $V(y) = V(y + 2a)$, $H_{\rm int}^{(n)}(y_n,\tau)$ must also be periodic in $y_n$ with the same 
period $2a$. Now add another particle to the right end of the chain. The total internal energy 
$H_{\rm int}^{(n+1)}(y_{n+1},\tau)$ of the resulting chain is related to $H_{\rm int}^{(n)}(y_n,\tau)$ as 
\begin{multline}
H_{\rm int}^{(n+1)}(y_{n+1},\tau)   = V(y_{n+1}) \\
 + \min_{y_n} \left \{ 
\frac{ \left (y_{n+1} - y_n - \mu \right )^2}{2\tau} + H_{\rm int}^{(n)}(y_n,\tau) \right \}. 
\label{eqn:Hy1}
\end{multline}

Recalling that $\mu$ is the equilibrium spacing of the springs, we can make a change of coordinates 
to positions relative to the endpoint of each unstretched spring as 
\begin{equation}
x_i = y_i - i \mu,  
\label{eqn:comoving}
\end{equation}
and Eq.~(\ref{eqn:Hy1}) becomes\cite{GC,CG} 
\begin{multline}
H_{\rm int}^{(n+1)}(x_{n+1},\tau) = V(x_{n+1} + (n+1)\mu) \\   
           + \min_{x_n} \left \{ 
\frac{ \left (x_{n+1} - x_n \right )^2}{2\tau} + H_{\rm int}^{(n)}(x_n,\tau) \right \}. \label{eqn:Hy2} 
\end{multline}

In this form the above equation closely resembles the problem presented in the previous section, 
Eqs.~(\ref{eqn:cat1}) and (\ref{eqn:cat2}), and we can evaluate the expression to be minimized on the 
RHS of Eq.~(\ref{eqn:Hy2}) via evolution of the inviscid Burgers equation, as follows:

Let 
\begin{equation}
u_n(x) =  \frac{\partial H_{\rm int}^{(n)}(x,\tau)}{\partial x},
\label{eqn:undef}
\end{equation}
then following the steps of the previous section and treating $t$ as a variable, we next define
\begin{equation}
H_{\rm int}(x,t) \equiv \min_{x^\prime} \left \{ 
\frac{1}{2t} \left (x - x^\prime \right )^2 + H_{\rm int}^{(n)}(x^\prime,\tau) \right \}.
\label{eqn:uxtau}
\end{equation}
Hence $u(x,t) \equiv \partial H_{\rm int}(x,t) / \partial x$ satisfies the inviscid Burgers equation 
\begin{equation}
u_t + u u_x = 0 \;\;\;\; \mbox{for} \;\;\;\;\; 0 \le t < \tau 
\label{eqn:udef}
\end{equation}
with  $u(x,0) = u_n(x)$, so that from Eqs.~(\ref{eqn:Hy2}) and (\ref{eqn:undef})
\begin{equation}
u_{n+1}(x) = V^\prime(x + (n+1)\mu ) + u(x,\tau).
\label{eqn:unp1def}
\end{equation}

Moreover, the position  $x_n$ of the $n^{\rm th}$ particle as a function of the position $x_{n+1}$ of particle $n+1$ is given by the characteristic mapping
\begin{equation}
x_{n+1} = x_n + u_n(x_n) \tau. 
\label{eqn:umap}
\end{equation}

The relations Eqs.~(\ref{eqn:undef}) -- (\ref{eqn:unp1def}) actually prescribe the evolution of $u$ under 
a forced Burgers equation. Defining  
\begin{equation}
\left.  u(x,t) \right \vert_{t = (n\tau)^{+}} = u_n(x),
\end{equation}
the evolution equations become
\begin{eqnarray}
 u_t + u u_x &=& 0 , \;\;\;\;\;\; n\tau \le t <  (n+1)\tau \label{eqn:evolutionStep}\\
 \left.\! u(x,t)\right \vert_{t = (n\tau)^{+}} &=& \left.\!\!u(x,t) \right \vert_{t = (n\tau)^{-}} + V^\prime(x + n\mu), \label{eqn:insertionStep}
\end{eqnarray}
which is equivalent to the periodically forced Burgers equation:
\begin{equation}
 u_t + u u_x = \sum_{n = 0}^{\infty} \delta(t - n\tau) \; V^\prime(x+n\mu),
\label{eqn:ForcedBurgers}
\end{equation}
with initial condition $u(x,0^{-}) = 0$ \cite{IC_Comm}. 

The flow of characteristics, Eq.~(\ref{eqn:umap}), under forced Burgers evolution implicitly defines the 
characteristic backwards map (also known as the Lagrange map), such that, given 
a final time $t_0$, for all $t \le t_0$
\begin{equation}
x(t) = x(t_0;t).
\label{eqn:Lagrange}
\end{equation}
For the configurations $\{x_i\}$ of the semi-infinite chain with the outmost particle $n$ being at $x_n$, 
this implies that for all $i \le n$ 
\begin{equation}
x_i = x(n\tau;i\tau), \;\;\;\;\;\;\mbox{with} \;\;\;\;\;\; x_n = x(n\tau;n\tau).
\label{eqn:xiLagr}
\end{equation}

We have therefore shown that a continuous one-parameter flow embodied by the 
Lagrange map Eq.~(\ref{eqn:Lagrange}), underlies the equilibrium configurations 
Eq.~(\ref{eqn:xiLagr}) of the discrete mass-spring system. The Lagrange map in turn 
is given by the backwards flow of the characteristic trajectories of the forced Burgers 
evolution Eq.~(\ref{eqn:ForcedBurgers}). Within this 
description the time-like evolution parameter $t$ is a material coordinate 
corresponding to the building up of springs by the continuous addition of 
material with elastic modulus $\mu/\tau$. 

Our derivation of the connection between the discrete particle configurations of a 
harmonic chain of particles and the characteristic flow of a forced Burgers equation 
has been based on an analysis of the forces acting on the particles. 
The connection between a general class of discrete minimization problems such 
as Eq.~(\ref{eqn:Hy1}) and certain one parameter flows was established first 
independently by Jausslin, Kreiss and Moser \cite{JKM} and 
E, Khanin, Mazel and Sinai \cite{EKMS}, using  
variational methods.
The connection with 
FK models in particular was developed further by E and Sobolevskii \cite{E,Sobolevskii}.

\subsection{Properties of the Characteristic Flow Pattern}
\label{charflowprop}

For the FK model the relevant results are as follows \cite{E,Sobolevskii,BecKhanin}: 
(i) To each asymptotic solution $u(x,t)$ there corresponds 
a flow pattern of characteristics $\gamma$ with  
$\gamma(t_0) = x_0$ which are traced {\em backwards} in time,  $t \in (-\infty,t_0]$, {\it cf.} 
Eqs.~(\ref{eqn:Lagrange}) and (\ref{eqn:xiLagr}). These characteristics 
$\gamma$ cannot cross each other and by construction, will never terminate in a shock. They are 
called {\it one-sided minimizers}. We should re-emphasize that by definition one-sided minimizers 
flow backwards in time. In the case of FK models they generate the lowest energy particle configurations of a semi-infinite chain with the outermost particle fixed at $x_0$. (ii) Among the one sided minimizers there exists a subset of minimizers 
that have the additional property that when traced {\it forward} in time, $t > t_0$, they never merge with 
a shock. These minimizers are the {\it global minimizers} and they correspond to the lowest energy 
configurations of the bi-infinite chain. As $t \rightarrow -\infty$, the one-sided 
minimizers converge to one of the global minimizers. (iii) Given a time $t$, the set of points $x_s$ such that minimizers immediately to its right and left ($x_s^+$ and $x_s^-$) converge to different global minimizers, constitute the locations of {\it global shocks}. Thus global shocks, if present, have the property 
that they can never disappear, as they separate the flows of one-sided minimizers that approach  different global minimizers. (iv) All minimizers associated with an asymptotic solution $u(x,t)$ have 
the same winding number $\ell/2a$, corresponding to the average spacing of particles of a configuration, 
{\it cf.} Eq.~(\ref{eqn:ell}). 
(v) Pinned particle configurations are characterized by the presence of 
shocks in the flow patterns. For rational $\ell/2a$ the flow pattern turns out to always contain shocks 
and the particle configurations will thus be pinned. For irrational $\ell/2a$, depending on the 
external potential, the asymptotic flow pattern may or may not contain any shocks. These cases correspond 
to pinned and sliding incommensurate configurations, respectively. 

For the particular FK model with piece-wise parabolic potential, which is the case we will be concerned here, the external forcing always contains a shock. Since a shock once present cannot disappear but at most will merge with another shock, shocks will be present in the flow pattern and the resulting particle 
configurations are pinned. Furthermore, almost all configurations (except a subset of measure 
zero) turn out to have rational winding numbers $\ell/2a = r/s$. 
The flow pattern is periodic in time $t$ with period $s\tau$\cite{E,BecKhanin} and thus global minimizers correspond to a periodic configuration of particles with period $s$. It is not hard to see that there must be $s$ global minimizers: At any time $t = n\tau$ their locations $x$ correspond to the $s$ distinct locations of particles in the periodic lowest energy configuration. Since starting from a time $t_0$, the backwards flow of one-sided minimizers 
converges to a global minimizer, these must converge to one of the $s$ global minimizers. Given 
that the configuration space $x$ is periodic (with the period of the external potential $2a$), the unit 
cell must contain $s$ global shocks separating the backwards flows of one-sided minimizers towards their associated global minimizer. In particular, the flow of the minimizers in the $xt$ plane will be 
confined to the interior of $s$ strips that are bounded by the trajectories of the global shocks 
and that each contain a global minimizer. The (backwards) flow of one-sided minimizers remains 
thus inside their respective strips and thereby converges towards the associated global minimizer.

\section{Burgers Description of the FK Model with Parabolic Potential}
\label{FKwithScalloped}

In this section we calculate the flow patterns of a    
FK model with a piece-wise parabolic potential. This case also 
corresponds to a strong pinning limit in which the external potential 
is so strong that particles in the lowest energy configuration  are 
confined to the vicinity of the minima of the potential wells, that can be 
treated  approximately as parabolic. The lowest 
energy configurations of this chain were calculated exactly by 
Aubry \cite{Aubry1983a}. The purpose of this section is 
to recover Aubry's solutions and to demonstrate how the forced Burgers 
evolution approach yields additional results and insights.

\begin{figure}[t]
 \centering \includegraphics[scale=0.7]{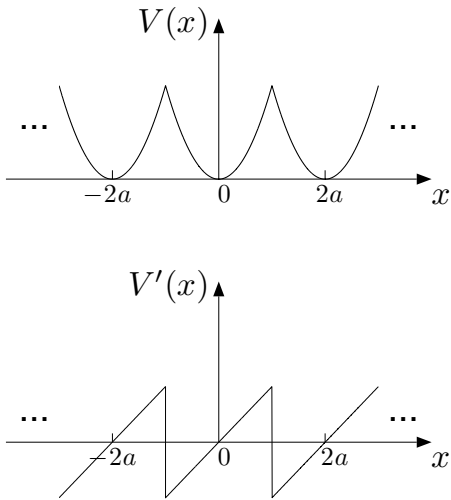}
\caption{Top: The piecewise parabolic potential $V(x)$, Eq.~(\ref{eqn:vpotagain}). Bottom: The corresponding piecewise linear profile $V^\prime(x)$.}\label{fig:pot_ramp}
\end{figure}

\subsection{Parameterization of the Burgers Profile and its Evolution}

The external potential is 
\begin{equation}
V(x)=\frac{1}{2}\lambda_{0}\left[x-2a\,{\rm Int}\!\left(\frac{x+a}{2a}\right)\right]^2, 
\label{eqn:vpotagain}
\end{equation}
where $2a$ and $\lambda_{0}>0$ are the period and the strength of the potential, respectively. 
We consider a unit cell that extends from $-a$ to $a$. The forcing of the Burgers equation, Eq.~(\ref{eqn:ForcedBurgers}), is given by $V^\prime(x)$ which is a series of ramps as shown in the bottom part of Fig.~\ref{fig:pot_ramp}. The $x$-intercepts correspond to the potential minima. The continuity of 
$V(x)$ across the boundaries of the unit cell further implies that the total area under the profile from $-a$ to $a$ is zero (area constraint).

\begin{figure}[t]
 \centering \includegraphics[scale=0.5]{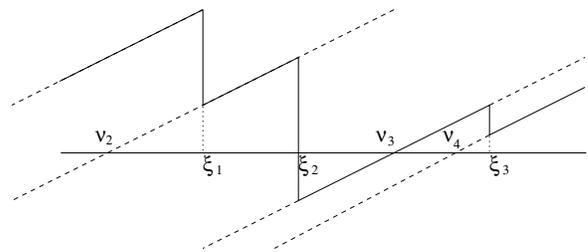}
\caption{The profile $u(x,t)$ and the parameterization of its linear segments.}\label{fig:representation}
\end{figure}

It is not difficult to see ({\it cf}. Appendix \ref{onBurgers}) that evolution under Eq.~(\ref{eqn:ForcedBurgers}) is such that for any time $t$, the Burgers profile within a unit cell consists of parallel straight line segments of slope $\lambda(t)$ terminated by shocks. A sample profile is shown in 
Fig.~\ref{fig:representation}. In what follows 
we will be making use of certain facts about the evolution of such profiles. The relevant results 
have been derived in Appendix \ref{ShockDetails}.

At any instant of its evolution, the profile of $u(x,t)$ is completely determined by a set of parameters \cite{Tatsumi}: Each line segment is part of an infinite line of slope $\lambda(t)$. Since the segments are confined between shocks, the position of the shocks determine the intervals that the segments occupy on their respective lines. Along with their slope, these lines are determined by their $x$-intercepts. 
Hence, if there are $\kappa$ shocks inside the unit cell, there are $\kappa+1$ segments. Numbering the segments within 
a unit cell from left to right as  $0$ to $\kappa$, we will denote the right terminations of each segment 
as $\xi^{(k)}$ and the corresponding $x$-intercept by $\nu^{(k)}$. Note that due to the periodicity 
across the 
unit cell, the intercept associated with the last segment is given as $\nu^{(\kappa)} = 2a + \nu^{(0)}$. Including the slope, $2\kappa + 1$ parameters are required to determine the profile $u(x,t)$ completely.

\begin{figure}[t]
 \centering
\includegraphics[scale=0.5]{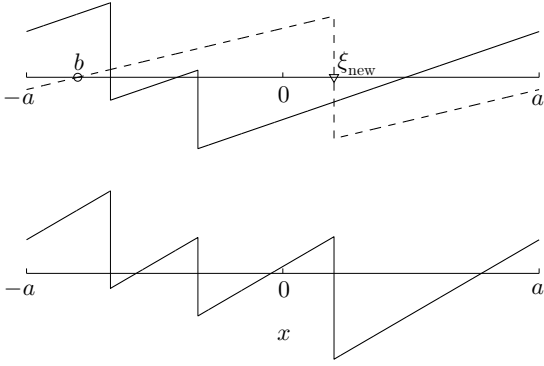}
\caption{Top: The profile $u$ and the profile $V^\prime$ to be superposed (dashed). Bottom: the resulting profile after the addition of the two profiles.}\label{fig:add_new_shock}
\end{figure}

The evolution of $u(x,t)$ under Eq.~(\ref{eqn:ForcedBurgers}) consists of two parts: the evolution step Eq.~(\ref{eqn:evolutionStep}) when a new spring is added to the end of the chain, and the particle insertion step Eq.~(\ref{eqn:insertionStep}), where a new shock is inserted, {\it cf.} Fig.~(\ref{fig:add_new_shock}). During the evolution step the slopes of the segments flatten according to Eq.~(\ref{eqn:lambdat}) and 
the shocks move and merge upon collision. Whenever a new shock is inserted, one of the linear segments of 
the profile will be split into two by the shock discontinuity of $V^\prime(x + n\mu)$ (unless it happens to coincide with the boundary of a segment), and the slopes of the profile will be incremented by $\lambda_0$, {\it cf.} Eq.~(\ref{eqn:vpotagain}).

Let us first consider the evolution of $\lambda(t)$. Denoting by
$\lambda^{\pm}_n$ the profile slope
just before and after addition of a particle at time $n\tau$, we see from Eq.~(\ref{eqn:lambdat}) that 
\begin{equation}
\lambda^{+}_{n+1} = \lambda_0 + \frac{\lambda^{+}_n}{ 1 + \lambda^{+}_n \tau}. 
\end{equation}
This recursion converges to a stable fixed point  \begin{equation}
\lambda^*_+=\frac{1}{2}\lambda_{0}\left(1+\sqrt{1+\frac{4}{\tau\lambda_{0}}}\right).
\label{lamstar}
\end{equation}
In what follows we will also need $\lambda^*_-/\lambda^*_+$, which, noting that  $\lambda^*_- = \lambda^*_+ - \lambda_0$, turns out to be
\begin{equation}
 \frac{\lambda^*_-}{\lambda^*_+} =  \frac{1}{1 + \lambda^*_+ \tau} = 
1 +\frac{\tau\lambda_{0}}{2}-\frac{\tau\lambda_{0}}{2}\sqrt{1+\frac{4}{\tau\lambda_{0}}} \equiv \eta,
\label{eqn:lambdaeta}
\end{equation}
coinciding with $\eta$ in\cite{Aubry1983a}. 
Asymptotically, the profile slopes right before and after a particle 
addition are thus given by $\lambda^{*}_{\pm}$. We will henceforth assume that sufficiently many 
particles have been added to the chain that the profile has reached its asymptotic slope. 

During the evolution step a segment $k$ disappears whenever shocks $\xi^{(k-1)}$ and $\xi^{(k)}$ merge. The evolution 
of a segment $k$ that survives the evolution step $n\tau < t < (n+1)\tau$ is given as, Eq.~(\ref{eqn:uline}),
\begin{equation}
 u^{(k)}(x,t) = \frac{\lambda^*_+}{1 + (t - n\tau)\lambda^*_+}\; (x - \nu^{(k)}) 
\end{equation}
for $\xi^{(k-1)}(t) < x < \xi^{(k)}(t)$.
Note that $\nu^{(k)}$ is constant for segments $k$ that survive the evolution step. The location of the intercept points $\nu^{(k)}$ will generally change during the particle insertion step, since besides creating a new segment, the slopes of all segments are augmented by $\lambda_0$, while the locations of the shocks $\xi^{(k)}$ already present remain unchanged. 

The mapping of $\nu^{(k)}$ during particle insertion can be worked out and is illustrated in Fig.~\ref{fig:add_new_shock}. 
Denote by $\nu^{\pm}$ the location of the segment intercepts before and after the particle addition, and 
let $b$ be the location of the intercept of the left segment of the new shock to be added (see Fig.~\ref{fig:add_new_shock}). The new shock thus is at $\xi^{({\rm new})} = b + a$ and the intercept to its right is at $b+2a$. One finds that 
\begin{equation}
\nu^{(k)+}-b^{(k)}=\eta \left(\nu^{(k)-}-b^{(k)}\right),
\label{nuupd}
\end{equation}
where $\eta$ is defined in Eq.~(\ref{eqn:lambdaeta}) and
\begin{equation}
b^{(k)} = \left \{ \begin{array}{ll} 
b, & \xi^{(k)} < \xi^{\rm new} \\
b + 2a, & \xi^{(k)} > \xi^{\rm new}.
\end{array} 
\right.  
\label{eqn:bi}
\end{equation}
Note that the segment intercepts, Eq.~(\ref{nuupd}), are \emph{attracted} toward their respective $b^{(k)}$'s,  since $\eta < 1$ unless $\lambda_0 = 0$. 

If, as will generally be the case, the new shock splits a particular segment $k$ into two, this will also create  an additional intercept, $\nu^{({\rm new})}$ that is given as
\begin{equation}
 \nu^{({\rm new})} = \eta \nu^{(k)-} + (1 - \eta) b.
\label{eqn:nunew}
\end{equation}
The ordered list of intercepts after insertion is thus 
\begin{equation}
\nu^{(0)+}, \nu^{(1)+}, \ldots, \nu^{(k-1)+}, \nu^{({\rm new})+}, \nu^{(k)+}, \ldots , \nu^{(\kappa)+}. 
\end{equation}
Since $\nu^{(k)+}$ remains constant for segments surviving the evolution step, we will henceforth 
drop the $+$ superscripts.

We consider next the evolution of the shocks. Denote by $\xi^{(k)}$ and $\nu^{(k)}$ the location of the shock and the corresponding segment intercept, respectively, {\em right after} a shock insertion. As we show 
in Appendix \ref{ShockDetails} the shocks move at constant velocity $v^{(k)}$ and we find, using  Eq.~(\ref{eqn:sxi}),
\begin{equation}
 v^{(k)}=\lambda^*_+ \left( \xi^{(k)} -\frac{\nu^{(k)} + \nu^{(k+1)}}{2}\right). 
\end{equation}
The final location $\xi^{(k)}_{\rm f}$ of a shock that survives the evolution step without merging 
with another shock is then found using Eq.~(\ref{eqn:lambdaeta}), as
\begin{equation}
\xi^{(k)}_{\rm f} = \frac{1}{\eta} \xi^{(k)} - \frac{1- \eta}{\eta} \; \frac{\nu^{(k)} + \nu^{(k+1)} }{2}.
\label{eqn:xievol}
\end{equation}

Finally, the parameter $b$ indicating the 
location of the zero intercept of $V^\prime(x + (n+1)\mu)$ in the co-moving frame, evolves according to 
\begin{equation}
b \rightarrow b - \mu. 
\label{eqn:b}
\end{equation}

Let subscripts $j$ denote the times $t = (j\tau)^+$ right after shock insertion. Together with the rules of how to handle colliding shocks given in Appendix \ref{ShockDetails}, we thus have a discrete dynamical system 
for the variables $\nu^{(k)}_j$ and $\xi^{(k)}_j$ that underlies the evolution of $u(x,t)$,  Eqs.~(\ref{eqn:eomfirst}) - (\ref{eqn:eomlast}).

For the FK model with piece-wise parabolic potentials, all lowest energy configurations are commensurate with an 
average spacing $\ell/2a = r/s$, where $r$ and $s$ are relatively prime integers. Correspondingly, the asymptotic behavior of $u(x,t)$ is periodic up to a shift in the sense that,  for all $x$ 
and $t$
\begin{equation}
u_*(x,t+s\tau) = u_*(x + s\mu,t). 
\label{eqn:uperiodic}
\end{equation}
Focusing on the moment right after a particle addition, this means in particular that there are 
$s$ different profiles, $u^+_*(x,0), u^+_*(x,\tau), u^+_*(x,2\tau), \ldots, u^+_*(x,(s-1)\tau)$ 
that turn out to  be related to the semi-infinite chain with its end particle located at one of the topologically distinct $s$ locations of the lowest energy configuration, as we will see shortly. 

Before proceeding with an analytical derivation of the steady-state profiles and the associated characteristic flow patterns, it is instructive to look at some steady-state solutions obtained by numerically evolving the profile parameters $\nu$ and $\xi$.

\subsection{Steady-State Profiles and Flow Patterns - Overview }

\begin{figure}[t]
\begin{center}
\includegraphics[scale=0.42]{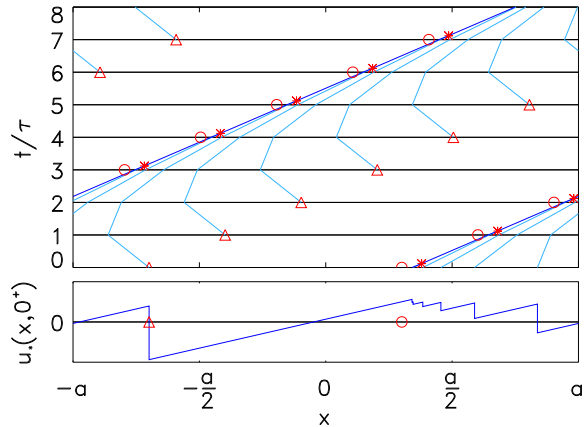}
\end{center}
\caption{(Color online) Top: Shock trajectories of the steady-state flow for parameter values $\mu/2a = 0.8495$ and $\lambda_0 = 0.4$. The period in this case is $\tau$, corresponding to $\ell/2a = 1$, and thus the 
profiles of $u(x,t)$ at $t = n\tau$ and $t = (n+1)\tau$ are equivalent up to shifts by $-\mu$, {\it cf.} Eq.~(\ref{eqn:uperiodic}).
At each time $t = n\tau$ a new shock is inserted due to the addition step and the insertion point 
is marked by a triangle. The red open circles show the locations of the well minima of the external potential (see text for further details). Bottom: Profile of $u(x,t)$ at 
$t = 0^+$.}
\label{fig:sample_one}
\end{figure}

Figure \ref{fig:sample_one} shows the shock trajectories associated with the steady-state flow of 
the forced Burgers equation with parameters $\mu/2a = 0.8495$ and $\lambda_0 = 0.4$, corresponding to 
 $\ell/2a = 1$. Since we will be only interested in steady-state flow patterns and not the transients, we 
have reset time to $t=0$. From Eq.~(\ref{eqn:uperiodic}) we see that 
the profiles $u(x,t)$ at $t = n\tau$ and $t = (n+1)\tau$ are equivalent up to shifts by $-\mu$.
At each time $t = n\tau$ a new shock is inserted due to the particle addition step and the insertion point 
is marked by a triangle. The bottom figure shows $u(x,t)$ at $t = 0^+$.

Recall that with respect to the co-moving frame the unit cell is moving by an amount of
$-\mu$ from one particle addition to the next, {\it cf.} Eq.~(\ref{eqn:comoving}). The boundaries of the unit cell 
coincide with the shock insertion locations, corresponding to the cusps of the external potential Eq.~(\ref{eqn:vpotagain}). They therefore also mark the location of the unit cell $y \in [-a,a)$ in the co-moving frame, corresponding to $y = \pm a$ in the unit cell coordinates. The locations of the minima of the external potential lie half-way between these two points at $y = 0$ in the cell reference frame and are marked by open circles. In terms of coordinates $y$ we have strict periodicity, $u_*(y,t+\tau) = u_*(y,t)$, {\it i.e.} without shift. 

Note how a newly inserted shock can survive subsequent insertions thereby creating a tree-like structure. 
We will refer to such structures as shock-trees. In fact, each inserted shock eventually collides with a previously inserted shock and the corresponding collision events have been indicated in the figure by red asterisks. Due to the periodicity  of the shock configurations, the life-time $\delta t_c$ of a newly inserted shock is constant. In the figure shown,  $\delta t_c$ is between $5\tau$ and $6 \tau$ implying 
that at any insertion time $t = (n\tau)^+$ there are 7 shocks including the newly inserted shock (two  
of the shocks are too close to be discerned, {\it see} bottom half of Fig.~\ref{fig:sample_one}). It is not hard to see that the time-periodicity of the profile also implies that all branches of the shock tree 
starting with a newly inserted shock are identical. They are merely nested on the tree due to their 
different creation times. 

Once the inserted shock merges with a previously inserted one, this 
shock proceeds to evolve until the next collision. One can think of the latter as an ever-present 
shock that keeps on absorbing newly inserted shocks, thereby constituting the trunk of the shock tree. 
This is the {\em global shock} introduced in section \ref{charflowprop}, while the shocks constituting the branches are the {\em secondary shocks} \cite{BecKhanin}. 

\begin{figure}[t!]
\begin{center}
\includegraphics[scale=0.47]{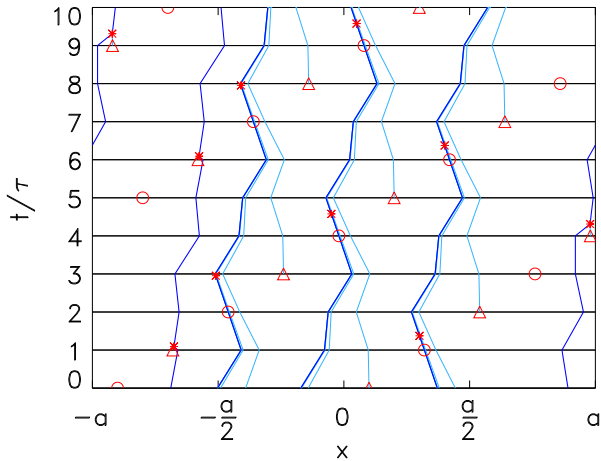}
 \end{center}
\caption{(Color online) Shock trajectories for $\mu/2a = 0.39$ and  
$\lambda_0 = 0.2$ at steady-state. The equilibrium spacing of the particles is $\ell/2a = 2/5$. 
Note that the figure contains $s = 5$ shock trees and that the tree patterns are periodic 
in $t = 5\tau$ up to an overall shift. For any shock tree, the secondary shocks (light blue) are either all to the left or the right of their global shock (dark blue). It turns out that at their 
respective insertion times the trees (from left to right) contain 2, 8, 7, 8 and 2 shocks, but most of these are too close to the global shock to be discerned.}
\label{fig:lro5traj}
\end{figure}

Figure \ref{fig:lro5traj} shows the shock trajectories of the steady state flow for $\ell/2a = 2/5$. There are $5$ shock trees. 
For each tree there is an 
insertion time at which a new shock is added. By the periodicity of the flow it follows that shocks are added to a given tree periodically (the period in this case being $5\tau$). 
Observe the ``feeding order'' of the trees. The immediate right neighbor of a tree on which a shock has been inserted is ``fed'' at the second subsequent insertion. As has been shown \cite{E,BecKhanin} ({\it see}  Section \ref{charflowprop}), the shock pattern for a configuration with  $\ell/2a = r/s$, with $r$ and $s$ irreducible integers, will always contain $s$ shock trees and the time periodicity of the pattern will be $t = s\tau$. Labeling the trees from left to right as $0, 1, 2, \ldots , s-1$, the feeding order of the trees turns out to be given by subtractions of $r$ mod $s$, as is proven in Appendix \ref{feeding-order}. From the periodicity $s\tau$ of the flow pattern it also follows that for a given shock tree all its secondary shocks are either to the left or right of its global shock. We will refer to such trees as left and right trees, respectively.
In other words, the time 
periodicity of $s\tau$ of the flow pattern and the presence of $s$ shock trees onto each of which a single secondary shock is inserted during the period $s\tau$, implies that a shock tree cannot 
have branches on both sides of its global shock. However, both types of trees can coexist, as seen in 
Figs.~\ref{fig:lro5traj} and \ref{fig:eql1o7}. 

We now turn to the flow pattern associated with shock trajectories at steady state. The lowest energy  configurations of a semi-infinite chain with its end point fixed are generated by the characteristic trajectories traced backwards in time according to Eq.~(\ref{eqn:xiLagr}). These trajectories are the {\em one-sided minimizers} \cite{BecKhanin}. Since the flow at steady state is time-periodic with period $s\tau$, it is sufficient to know the backwards flow for times $t \in [0,s\tau)$ and for all $x \in [-a,a)$, generating a map that can be iterated.

We expect that as the characteristics are traced backwards in time, corresponding to particles deeper and deeper inside the chain, the effect of the location of the particle at its end point will diminish. As we will show, the effect of the boundary decays as $\eta^i$. This means that as the characteristics are traced further back, the characteristic trajectory will approach a limiting cycle (due to the periodicity of the steady-state flow pattern). Since the effect of the boundary will have vanished, the particle configuration generated by the limit-cycle must be the lowest energy configuration of the bi-infinite chain. This confirms the remarks in Section \ref{charflowprop}, namely that one-sided minimizers when traced backwards in time will  
converge to a global minimizer. Equivalently, the global minimizers when traced forward and backwards 
in time, generate the lowest energy configuration of the bi-infinite chain.

\begin{figure}[t]
\begin{center}
\includegraphics[scale=0.47]{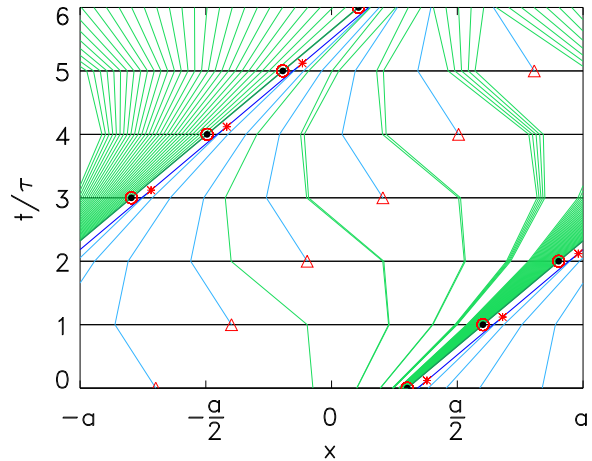}
\end{center}
\caption{Shock trajectories and backwards flow of the characteristics (minimizers) for $\mu/2a = 0.8495$ and  
$\lambda_0 = 0.4$ at steady-state. The average spacing of the particles is $\ell/2a = 1/1$. 
The annotation of the shocks and unit cell boundaries are as in Fig.~\ref{fig:sample_one}. The trajectories in green denote {\em one-sided minimizers}, whereas the dark green trajectory is the {\em global minimizer} that generates the lowest energy configuration in the co-moving frame. The locations corresponding 
to the positions of the particles in this configuration have been marked on the global minimizer by black solid circles.}
\label{fig:eql1o1}
\end{figure}

\begin{figure}[t]
\begin{center}
\includegraphics[scale=0.47]{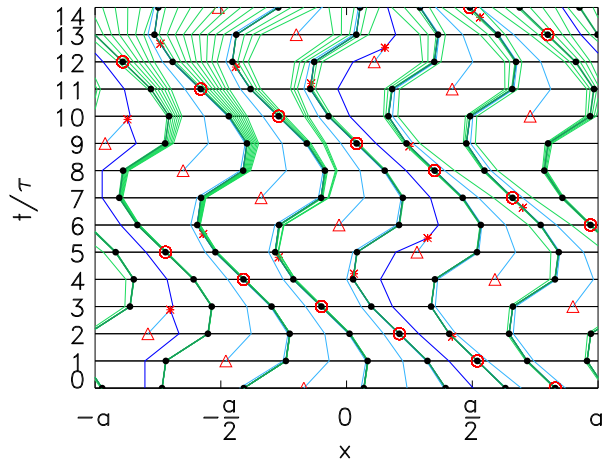}
\end{center}
\caption{(Color online) Shock trajectories and backwards flow of the minimizers for  $\mu/2a = 0.155$ and  $\lambda_0 = 0.2$ at steady-state, corresponding to $\ell/2a = 1/7$. The flow pattern is marked 
by the presence of $s = 7$ shock trees. The annotation is as in Fig.~\ref{fig:eql1o1}. Note that for 
some of the trees the main shocks (dark blue) are obscured by the global minimizers (dark green), 
since they are very close to each other (see text for further details).}
\label{fig:eql1o7}
\end{figure}

Figure \ref{fig:eql1o1} shows the flow pattern associated with Fig.~\ref{fig:sample_one}. The one-sided minimizers correspond to green lines, while the global minimizer is indicated by a darker green line. 
We have $\ell/2a = 1/1$, meaning that each potential well contains one particle. The location 
of the particles of the lowest energy configuration in the co-moving frame are shown as black solid circles and they necessarily lie on the global minimizer.  Furthermore, they turn out to coincide with the 
locations of the minima of the potential wells (red open circles). Figure \ref{fig:eql1o7} shows the flow pattern for $\ell/2a = 1/7$. Note how the one-sided minimizers 
when traced back in time flow onto one of the seven global minimizers. Again, the region between any two 
shock trees contains precisely one global minimizer and the corresponding particle configurations in the 
fixed frame are all equivalent up to an overall cyclic permutation.

\subsection{The Fundamental Shock Tree}
\label{FundamentalShockTree}

Given a configuration with $\ell/2a = r/s$, the corresponding 
flow pattern will contain $s$ shock trees and also $s$ global shocks. The global shocks 
have the property that they do not disappear. The secondary shocks constitute the 
branches of the shock tree that will eventually merge with the global shocks. 
The presence of a gap region in between shock trees that is bounded on each side by a shock, 
implies that the corresponding segment of $u(x,t)$ will never disappear, since the shocks 
at its boundaries will never merge with each other. 
We will refer to these segments as {\em global segments}. Likewise, the intercepts associated with such 
segments will never disappear and we will refer to them as the {\em global intercepts}.

One note of caution is in order. Recall that our convention has been to
identify the intercept of each continuous segment of $u(x,t)$ with 
the same label as the shock that bounds it on the right. With this convention 
one must be careful since the global shock bounding a global
segment may fall to the left of the segment, and therefore the global segment and 
the global shock may not have the same label. 
This occurs for a left tree: By definition, the rightmost shock in a left tree
is its global shock. The segment bounded on the right by this shock is
however not a global segment, since this segment and its intercept will 
disappear when the secondary shock bounding the segment from its left merges 
into the global shock. It is not difficult to see that for a left tree  the 
segment bounded on the left by the global shock is a global segment. Thus its label 
will be that of the shock lying immediately to the right of the global shock. 
In general this shock will belong to another tree \cite{oneshocktree}. 
In the case of a right tree such a situation does not occur. The segment 
associated with the global shock is global as well and therefore carries the same 
label.

The segments of $u(x,t)$ associated with the global intercepts are by construction the segments of 
$u$ that span the gap region in between two neighboring shock trees. This region contains the 
global minimizer and moreover governs also the backwards flow of one-sided minimizers in its 
vicinity. 
Except for the case where a right tree is immediately to the right of a left tree, the global region 
will be bounded on at least one side by secondary shocks. Thus whenever a new secondary shock is 
inserted, the corresponding global segment will be intersected, 
spawning off two intercepts: one that remains global and one that is associated with the 
secondary shock. We will refer to the latter as {\it secondary} or {\it local intercept}.  An example is shown in the right panel of Fig.~\ref{fig:xinu_example}. 
Here the flow pattern contains a single shock tree that is of right type. Thus the global 
region is bounded by the two sides of the tree, which is equivalent to being bounded by two right trees. 
The global intercepts are shown in red, 
while the secondary intercepts are in green. The bifurcation into two intercepts 
occurs at $t = n\tau$ but for clarity has been offset in time by a small amount. 
As can be seen, the segment associated with the global region is bisected by successive shock insertions. 
The corresponding global intercept spawns off local intercepts that evolve on their own. The local 
intercepts are terminated when the corresponding secondary shock merges with the global shock.

The time periodicity of the steady state flow, 
Eq.~(\ref{eqn:uperiodic}),  implies that the number of shocks and  
intercepts is conserved across a period. Since a shock insertion always adds a new shock and 
a corresponding intercept, while a shock collision removes a shock and an intercept, it follows 
that during the period $s\tau$ the number of shocks inserted must equal to the number 
of shocks that merged with the global shocks.  

\subsection{The Global Intercepts}

The global intercepts $\nu$ can now be obtained as follows. 
By definition, global intercepts do not disappear and their locations remain constant 
during Burgers evolution. However during shock insertions they are remapped according to 
Eqs.~(\ref{eqn:eomfirst}), (\ref{eqn:eomsecond}) and (\ref{eqn:eomthird}). This mapping 
in turn depends on whether the shock associated with $\nu$ is to the left or right of the 
newly inserted shock. Thus we first have to determine the sequence $b^{(k)}_j$ of 
Eq.~(\ref{eqn:eomthird}) governing the evolution of the global intercepts. This is related 
to the ``feeding order'' in which new shocks are inserted into the trees
and the details are given in Appendix \ref{feeding-order}. For the segment associated with the global intercept 
of a shock tree the result is,   
\begin{equation}
b_j = -j\mu + 2a \; {\rm Int}\left [ (j+\delta)\frac{r}{s} \right ],
\label{eqn:bi2}
\end{equation}
where henceforth we will omit the superscript $(k)$, whenever the segment in question is global. The 
integers $\delta = 0, 1, \ldots, s-1$ are each associated with one of the $s$ shock trees. They are related to the relative time lags of each shock tree from their next shock insertion.

By definition, the global intercepts  
survive all evolution steps. This turns Eq.~(\ref{eqn:eomsecond}) together with Eq.~(\ref{eqn:bi2}) into 
a recursion relation that can be solved. For our purposes it is more convenient to 
consider the intercepts in the unit cell coordinates defined as follows:
\begin{equation}
\tilde{\nu}_j \equiv \nu_j - b_j,  
\end{equation}
which defines the location of the intercept relative to the location of the minimum of the 
potential well, so that $\tilde{\nu}_j \in [-a,a)$. As will be shown shortly, without loss of 
generality we will consider the shock tree for which $\delta = 0$. Then, the above definition along with 
Eqs.~(\ref{eqn:eomsecond}) and (\ref{eqn:bi2}) yields the following recursion
\begin{equation}
\tilde{\nu}_{j+1} = \eta  \tilde{\nu}_j + \eta \mu - 2a \eta \chi_j, 
\end{equation}
where $\chi_j \equiv (b_{j+1} - b_j  + \mu)/2a $ is given as 
\begin{equation}
\chi_j = {\rm Int}\left ((j+1)\frac{r}{s} \right ) - {\rm Int}\left (j\frac{r}{s} \right ).
\label{eqn:chidef}
\end{equation}
Note that $\chi_j$ is periodic, $\chi_{j+s} = \chi_j$, as well as $\chi_0 = 0$ and $\chi_{s-1} = 1$ 
(except for the case $r=s=1$, for which $\chi_0 = 1$).
In the cell coordinates, due to the time periodicity of the profile we also have $\tilde{\nu}_{j+s} = \tilde{\nu}_j$. Upon solving the recursion  we find
\begin{equation}
\tilde{\nu}_{0} = \mu \frac{\eta}{1-\eta} - \frac{2a}{1 - \eta^s} \; \sum_{i = 0}^{s-1} \eta^{s-i}\chi_i. 
\label{eqn:nueq}
\end{equation}
The remaining $\tilde{\nu}_j$, for $j = 1, 2, \ldots, s-1$ can be then found as
\begin{eqnarray}
\tilde{\nu}_j &=& \mu \frac{\eta}{1-\eta} \\
&-& 2a \left \{ 
\frac{\eta^j}{1 - \eta^s} \; \sum_{i = 0}^{s-1} \eta^{s-i}\chi_i + \sum_{i = 0}^{j-1} \eta^{j-i}\chi_i
\right \}. \nonumber 
\label{eqn:nujeq}
\end{eqnarray}

From  the definition of $\chi_j$, Eq.~(\ref{eqn:chidef}), and its relation to $b_j$, Eq.~(\ref{eqn:bi2}), it is evident that a non-zero $\delta$ will induce 
a cyclic shift  of $\chi$ by an amount of $\delta$. We thus define 
\begin{equation}
\chi_i^\delta = \chi_{(i + \delta) \; {\rm mod}\; s}, 
\end{equation}
so that for $\delta > 0$ this is a left shift. Using Eq.~(\ref{eqn:nueq}) to define a function 
$\tilde{\nu}[\chi]$ such that $\tilde{\nu}_0 \equiv \tilde{\nu}[\chi]$, the periodicity of $\tilde{\nu}_j$ 
implies 
\begin{equation}
\tilde{\nu}_j = \tilde{\nu} \left [\chi^{j} \right ].
\label{eqn:nudelta}
\end{equation}

At any time $t = n\tau$, the global intercepts associated with each of the $s$  shock trees only differ by 
the time of last insertion of a shock, as captured by the distinct values of $\delta$. Thus from  
Eq.~(\ref{eqn:nudelta}) we see that the set of global intercepts at any given time must also 
coincide with the set $\{ \tilde{\nu}_j\}$ (up to a cyclic permutation of its elements). 
In other words, $\{ \tilde{\nu}_j\}$ not only corresponds to the periodic sequence of global intercepts 
associated with a given shock tree  during its time evolution, it also corresponds to the set of 
all global intercepts associated with the $s$ shock trees at any given time $t = n\tau$.  
Thus with respect to their global intercepts all shock 
trees are alike, differing only in their respective shock insertion times. 

In fact, it can be shown 
that essentially the same holds true for all intercepts associated with the shocks as well as the secondary shocks themselves:  If we label the secondary shocks and their
corresponding intercepts at a time $t = j\tau$ on a given tree by a superscript $k$, then it turns out there is an infinite sequence $\tilde{\xi}_j^{k}$, $\tilde{\nu}_j^{k}$, with $j = 0, 1, \ldots, s-1$, and $k = 0, 1, 2, \ldots$, such that for each shock tree $\alpha$ with $\kappa_\alpha + 1$ secondary shocks, the actual secondary 
shocks and intercepts are subsequences terminated at $k = \kappa_\alpha$. This reduces the problem of obtaining the steady state shock pattern to finding the $s$ global shocks and their $\kappa$ values. The calculations are rather involved and will be carried out elsewhere \cite{MM}. Instead, here we will restrict ourselves to $s=1$ which is a special case of the above
and already contains most of the relevant features of the general case.
On the other hand, the global minimizers and thus the lowest energy configurations can be 
determined from the global intercepts alone, which we have just found for all $r$ and $s$. 
We will carry this out next.  

\subsection{Lowest Energy Configurations}

We turn first to the evolution of characteristics inside the global segments. 
As mentioned before, the characteristics $x(t)$ associated with the lowest energy configuration 
are the global minimizers and have the property that they are periodic up to a 
shift,  
\begin{equation}
x(t + s\tau) = x(t) - s\mu. 
\end{equation}
Our goal will be therefore to write the 
flow equation of characteristics inside the global segment and then impose the 
periodicity condition. Since we are interested in a periodic solution this calculation can 
be done forward or backwards in time, but it turns out to be more convenient to consider 
the forward flow of characteristics. Letting $x_j = x(j\tau)$, the characteristic equation for any $x_j$ 
within the global segment is given by
\begin{eqnarray}
x_{j+1} &=& x_{j} + \lambda^*_+\tau \left (x_{j} - \nu_{j} \right ) \nonumber \\
    &=& \frac{1}{\eta} x_{j} - \frac{1-\eta}{\eta} \nu_{j}.
\end{eqnarray}
Introduce the unit cell coordinates $\tilde{y}_j$ as
\begin{equation}
x_j = \tilde{y}_j + b_{j}
\end{equation}
so that, using Eqs~(\ref{eqn:bi2}) and (\ref{eqn:chidef}), the recursion for $\tilde{y}$ becomes
\begin{equation}
\tilde{y}_{j+1} = \frac{1}{\eta} \tilde{y}_j - \frac{1-\eta}{\eta} \tilde{\nu}_{j} 
+   \mu - 2a \chi_{j} , 
\end{equation}
which has the solution
\begin{eqnarray}
\tilde{y}_j &=& \frac{1}{\eta^j} \tilde{y}_0 - \frac{1-\eta}{\eta}\sum_{i=0}^{j-1} \eta^{j-i-1}\tilde{\nu}_i
\nonumber \\ 
&-& 2a \sum_{i=0}^{j-1} \eta^{j-i-1} \chi_i + \mu\; \frac{1-\eta^j}{1-\eta}.
\end{eqnarray}

Substituting the expression for $\tilde{\nu}_j$, Eqs.~(\ref{eqn:nueq}) and (\ref{eqn:nujeq}), imposing 
the periodicity condition $\tilde{y}_0 = \tilde{y}_s$, one finds that  
\begin{equation}
\tilde{y}_0 = \frac{2a\eta}{(1+\eta)(1-\eta^s)} \; 
\sum_{k=0}^{s-1} \left ( \eta^k  - \eta^{s-1 - k} \right ) \chi_k,
\label{eqn:x0}
\end{equation}
which can also be rewritten as
\begin{equation}
\tilde{y}_0 = \frac{2a\eta}{(1+\eta)(1-\eta^s)} \; 
\sum_{k=0}^{s-1} \eta^k \left ( \chi_k - \chi_{s-1 - k} \right ).
\label{eqn:xzero}
\end{equation}

We can calculate again the remaining $\tilde{y}_j$ from the cyclic permutation property.
With $\chi^\delta$ as given above,  we define $\tilde{y}[\chi]$ such that using Eq.~(\ref{eqn:x0}) we 
have $\tilde{y}_0 = \tilde{y}[\chi]$.  
Then 
\begin{equation}
\tilde{y}_j = \tilde{y}[\chi^{-j}], 
\label{eqn:xchishift}
\end{equation}
or explicitly, 
\begin{eqnarray}
\tilde{y}_j &=& \frac{2a\eta }{(1+\eta)(1-\eta^s)} \;  \\
&\times& \; \sum_{k=0}^{s-1} \eta^k \left \{
\chi_{(k + j) \; {\rm mod} \; s} - \chi_{(s - 1 + j - k) \; {\rm mod} \; s} \right \} \nonumber
\label{eqn:xj}
\end{eqnarray}

It can be shown that in terms of the 
hull-function $f(x)$ of Aubry's solution Eq.~(\ref{eqn:fgeq}), the above equation upon substitution of Eq.~(\ref{eqn:chidef}) reduces to
\begin{equation}
\tilde{y}_j = f \left (j\frac{r}{s}\; 2a \right ) - 2a \; {\rm Int} \left ( j\frac{r}{s} \right ) \equiv g\left (j\frac{r}{s} \; 2a \right ), 
\end{equation}
with $g(x+2a) = g(x)$, from which Aubry's result\cite{Aubry1983a} follows under the identification
\begin{equation}
y_j = \tilde{y}_j + 2a \; {\rm Int}\left ( j \frac{r}{s} \right ).
\end{equation}

Once the lowest energy configuration has been found, the mode-locking intervals of $\mu$ over which 
a given average spacing $\ell/2a = r/s$ is a lowest energy configuration as well as other properties 
pertaining to this configuration such as the Peierls-Nabarro barrier are readily obtained \cite{Aubry1983a}.

In other words, these are properties that can be obtained from the global minimizers alone. The global 
region is terminated by shocks, and thus as long as the location of the shocks are not known, we 
do not know the extent of the global region.
It is therefore impossible to know which one-sided minimizers will flow towards which global minimizer. 
We thus turn next to the calculation of the flow pattern which will also allow us to 
understand further what happens at the boundaries of the mode-locking intervals.   
We restrict the analysis to the case $s=1$, which already contains the 
relevant features of the general case. 

\subsection{Steady-state Flow Pattern for $s=1$}

\begin{figure*}[t]
\begin{center}
\includegraphics[scale=0.95]{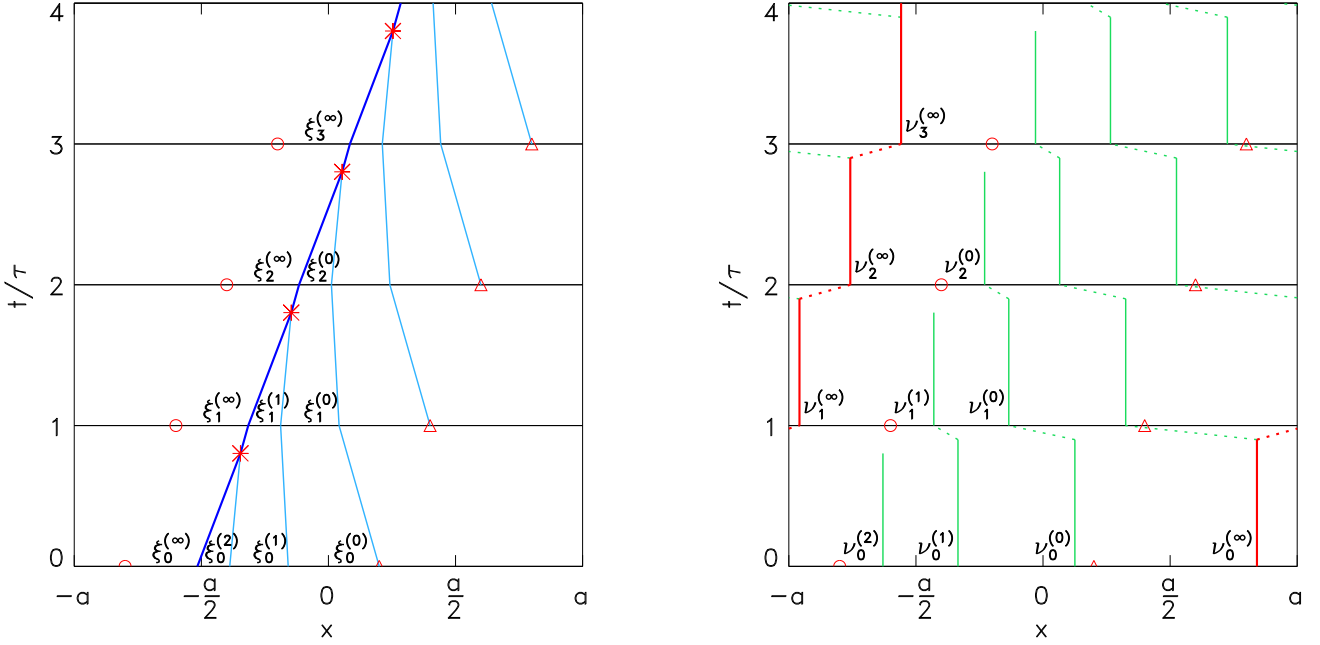}
\end{center}
\caption{(Color online) Steady-state shock trajectories (left) and evolution of intercepts $\nu$ (right) for  $\mu/2a = 0.9$ and  $\lambda_0 = 0.2$, corresponding to $\ell/2a = 1/1$. There are $\kappa + 2 = 4$ shocks right after insertion.  Left: The shock tree is of right type. The global shock is shown in dark blue, while secondary shocks are light blue.  
Right: Global intercepts are shown 
in red, while local intercepts are shown in green. At a shock insertion, a segment 
is bisected giving rise to a bifurcation of the $\nu$ intercept associated with that 
segment. The resulting pair of intercepts has been linked to the parent intercept 
by dotted lines and the times $t = n\tau$ at which the bifurcation occurs has been offset 
to a slightly earlier time for clarity. See text for details on the labeling of shocks and intercepts.}
\label{fig:xinu_example}
\end{figure*}

The steady state flow pattern contains a single shock tree and 
we will consider only the case $s = r = 1$, for which $\chi_0 = 1$ \cite{note}. 
As we have pointed out before, 
for a single shock tree the global segment is always intersected by a shock insertion. As can be seen from 
the right panel of Fig.~\ref{fig:xinu_example}, the particle insertion steps cause a bifurcation 
of the global intercept $\nu$ into a left and right intercept. For the case $\chi_0 = 1$, the 
equation for $b_j$ that governs the evolution of the global intercept, Eq.~(\ref{eqn:bi2}), becomes 
\begin{equation}
 b_j = -j\mu + 2aj, 
\end{equation}
which implies that the global intercept $\nu$ remains on the right half of the bifurcation, 
{\it cf.}~Eq.~(\ref{eqn:bi}). This is 
equivalent to saying that the corresponding shock tree is a right tree. 
From Eq.~(\ref{eqn:nueq}) we find
\begin{equation}
 \tilde{\nu}_0 = - (2a - \mu)\; \frac{\eta}{1-\eta}.
\end{equation}
The left branch $\nu^{\rm L}_j$ of the bifurcation must obey the recursion for $\nu$, Eq.~(\ref{eqn:eomsecond}), 
with $b^{\rm L}_j = b_j - 2a$ and we thus have
\begin{equation}
\nu^{\rm L}_{j+1} =  \eta \nu^{\rm L}_j + (1-\eta) (b_j - 2a).
\end{equation}

With reference to the right panel of Fig.~\ref{fig:xinu_example}, consider the time  $t  = 0^{-}$ at 
which the global intercept bifurcates. Denote the common ancestor at $t = -\tau$ as $\nu^{(0)}_{-1}$ so 
that $\nu^{\rm L}_{-1} = \nu_0 + 2a$.
Following the secondary intercept for times $j\tau$ with $j \ge 0$, moving into the cell 
coordinates and solving the recursion for 
$\tilde{\nu}^{\rm L}_{j}$ we find that 
\begin{equation}
\tilde{\nu}^{\rm L}_{j} = - \left ( 2a - \mu \right ) \; \frac{\eta}{1-\eta} + 2a \eta^{j+1}. 
\end{equation}
For reasons that will be apparent soon, we label the intercepts of a shock tree at time $t = 0$ as  $\tilde{\nu}^{(k)}_{0}$, with $k=0, \ldots \kappa$ so that the number of secondary intercepts and shocks is 
each $\kappa + 1$, as shown in the right panel of Fig.~\ref{fig:xinu_example}. 
With this labeling $\xi^{(0)}_0$ denotes the newly inserted shock at $t = 0$. 
From the time periodicity $\tau$ of the flow pattern it then follows that  
\begin{equation}
 \tilde{\nu}^{(k)}_{0} = \tilde{\nu}^{\rm L}_{k} \equiv \tilde{\nu}^{(0)}_{k}  
\end{equation}
and thus
\begin{equation}
\tilde{\nu}^{(k)}_{0} = - \left ( 2a - \mu \right ) \; \frac{\eta}{1-\eta} + 2a  \eta^{k+1} . 
\label{eqn:nutilde_k}
\end{equation}
Therefore in the unit cell coordinates, the time evolution of a secondary intercept when projected back onto the set of intercepts at $t = 0$ is simply a shift $\tilde{\nu}^{(k)}_{0} \rightarrow \tilde{\nu}^{(k+1)}_{0}$. 
This can be clearly seen in the right panel of Fig.~\ref{fig:xinu_example}, where the open circles mark 
the origin of the unit cell. 

For a steady state flow pattern with  $\kappa + 2$ intercepts, the global 
intercept must map into itself, the secondary intercept $k = \kappa$ must 
disappear upon further evolution and a new intercept is created at shock insertion. 
Thus the number of secondary intercepts remains the same from one shock insertion to the next. Since intercepts can only disappear if their corresponding 
shocks merge with other shocks, a steady state of the shift pattern  
requires that the secondary shock associated with the segment $k = \kappa$ collides with the global 
shock. 
We have already found the location of the global intercept  $\tilde{\nu}_{0}$. In fact, note 
that from Eq.~(\ref{eqn:nutilde_k}) we have that $\tilde{\nu}^{(\infty)}_{0} = \tilde{\nu}_{0}$. 
The shift 
$\tilde{\nu}^{(k)}_{0} \rightarrow \tilde{\nu}^{(k+1)}_{0}$ of the local intercepts under time evolutions then  
suggests to label the global intercepts and shock as $\tilde{\nu}^{(\infty)}_{0}$ 
and $\tilde{\xi}^{(\infty)}_{0}$, respectively.  The labeling of intercepts and their corresponding 
shocks is shown in  Fig.~\ref{fig:xinu_example}.

Thus given the steady 
state profile in the cell coordinate system, the ordering of the corresponding 
intercepts is
\begin{equation}
\tilde{\nu}^{(\infty)}_{0}, \tilde{\nu}^{(\kappa)}_{0}, \tilde{\nu}^{(\kappa-1)}_{0}, \ldots, 
\tilde{\nu}^{(1)}_{0}, 
\tilde{\nu}^{(0)}_{0}
\end{equation}
with 
\begin{equation}
 \tilde{\nu}^{(\infty)}_{0} = \tilde{\nu}_{0}.
\end{equation}
The number of shocks constituting a  shock tree, $\kappa +2$, at a given insertion time is directly related to the 
lifetime $\delta t_c$ of a shock from its insertion to its absorption by the global shock as $\kappa \equiv {\rm Int} (\delta t_c/\tau)$. 

The locations of the secondary shocks can now be found as follows. Consider time $t = 0$ and let  
$\xi^{(0)}_0, \xi^{(1)}_0, \ldots , \xi^{(\kappa)}_0$ denote the initial locations of the secondary shocks 
with the labeling in correspondence with that of the associated secondary intercepts, as shown in the left panel of Fig.~\ref{fig:xinu_example}. 
The subsequent positions at time $t = j\tau$ are obtained from Eq.~(\ref{eqn:eomforth}). In terms of  
the cell coordinates $\tilde{\xi}^{(k)}_j = \xi^{(k)}_j - b_j$, we find
\begin{equation}
\tilde{\xi}^{(k)}_{j+1} = \frac{1}{\eta}\; \tilde{\xi}^{(k)}_j - \frac{1-\eta}{\eta} \; \frac{\tilde{\nu}^{(k)}_j + \tilde{\nu}^{(k-1)}_j}{2}.
\label{eqn:xirec}
\end{equation}
where $\tilde{\nu}^{(-1)}_j = \tilde{\nu}^{(\infty)}_j + 2a$ which compensates for the wrapping around the cell boundary for $k=0$.
The shift property of the secondary intercepts necessarily applies to their associated secondary 
shocks as well,  so that $\tilde{\xi}^{(k)}_{j+1} = \tilde{\xi}^{(k+1)}_{j}$. 
Furthermore, by the shift property $\tilde{\xi}^{(k)}_{j} = \tilde{\xi}^{(k+j)}_{0}$ 
for all $k,j k+j \le \kappa$, so that without loss of generality we  can restrict ourselves to $j = 0$. 
For a right tree, it is convenient to let the location of the newly inserted shock in the cell 
coordinates be $\tilde{\xi}^{(0)}_0 = +a$ and one finds 
\begin{equation}
\tilde{\xi}^{(k)}_{0} =  a \eta^k\;\;\;\;\; k = 0, 1, 2, \ldots, \kappa. 
\label{eqn:xisec}
\end{equation}

The global shock $\tilde{\xi}^{(\infty)}_0$ and $\kappa$ are still undetermined, since so far  
we have not dealt with the collisions that must necessarily occur. We have obtained  
all intercepts as well as the positions of the secondary shocks.
 The steady-state shift motion of 
the intercepts described above implies that secondary shocks should not collide with each other 
during their time evolution. We will now 
show this explicitly by determining the time required for two adjacent 
secondary shocks to collide. 
Due to the time periodicity  $\tau$ it 
is sufficient to do the calculation at $t = 0^+$. The velocity of a secondary shock $\tilde{\xi}^{(k)}_{0}$ 
is given as
\begin{equation}
v^{(k)} = \lambda^*_+ \left ( \tilde{\xi}^{(k)}_{0} - \frac{\tilde{\nu}^{(k)}_{0} + \tilde{\nu}^{(k-1)}_{0} }{2} \right )
\label{eqn:vk1}
\end{equation}
for $k = 0, 1, 2, \ldots, \kappa$, where due to wrapping around the unit cell boundary we have again $\tilde{\nu}^{(-1)}_0 \equiv \tilde{\nu}^{(\infty)}_0 + 2a$. Letting $t^{(k)}_c$ be the time of collision between shocks 
$k$ and $k+1$, and noting from Eq.~(\ref{eqn:lambdaeta}) that $\lambda^*_+ \tau = (1-\eta)/\eta$, 
\begin{equation}
\frac{t^{(k)}_c}{\tau} = - \; \frac{1-\eta}{\eta}\; 
\frac{\tilde{\xi}^{(k)}_{0} - \tilde{\xi}^{(k+1)}_{0}}{v^{(k)} - v^{(k+1)}}.  
\label{eqn:tk1}
\end{equation}
We find for $k < \kappa$, that $t^{(k)}_c/\tau= 1/(1-\eta)$ and thus all secondary shocks will collide simultaneously, if at all. However, since $\eta \in [0,1]$, $t_c \ge \tau$, 
two secondary shocks cannot collide during the time evolution $0 < t < \tau$ (except for the 
case $\eta = 0$, corresponding to an infinitely strong external potential which we will ignore). Since the flow pattern has time periodicity $\tau$, this moreover means that they can never collide. 
Thus the only collision possible is between the global shock and its adjacent secondary shock. 

The global shock location can be determined by use of the area constraint,  
\begin{equation}
\int_{-a}^{a} u(x,t) \; {\rm d}x = 0, 
\end{equation}
which follows from the continuity of the internal energy $H_{\rm int}(x+2a,\tau) = H_{\rm int}(x,\tau)$.
We find
\begin{equation}
\tilde{\xi}^{(\infty)}_0 = \frac{1}{\eta^\kappa} \; \left [
\frac{a}{1+\eta} - \frac{2a - \mu}{1-\eta} \right ] + a \frac{\eta^{\kappa + 1}}{1 + \eta}. 
\label{eqn:xiglobal}
\end{equation}

A few points are worth noting. While the locations of the secondary shocks in the cell coordinates, 
Eq.~(\ref{eqn:xisec}) are independent of $\mu$, the location of the global shock does depend on $\mu$. 
Moreover, the factor $\eta^{-\kappa}$ in front of the term in rectangular brackets will diverge 
as $\kappa \rightarrow \infty$ unless the expression in the brackets vanishes sufficiently fast, which 
for each $\eta$ puts a constraint on $\mu$ as a function of $\kappa$, establishing thereby the region of $\mu$ and $\eta$ values for which a steady-state flow pattern with $s = r = 1$ can be obtained. 

We next look at the conditions under which the global shock $\tilde{\xi}^{(\infty)}_0$ can collide 
with its adjacent secondary shock $\tilde{\xi}^{(\kappa)}_0$, as required by the flow properties 
of the steady state solution. Denoting by $t^{(\kappa)}_c$ the time of collision, the requirement is 
\begin{equation}
0 < \frac{t^{(\kappa)}_c}{\tau} < 1.
\label{eqn:tineq}
\end{equation}
Next, $t^{(\kappa)}_c$ is found using Eqs.~(\ref{eqn:vk1}) and (\ref{eqn:tk1}) with 
$\tilde{\nu}^{(\kappa + 1)}_0 \equiv \tilde{\nu}^{(\infty)}_0$ and 
$\tilde{\xi}^{(\kappa + 1)}_0 \equiv \tilde{\xi}^{(\infty)}_0$. 
Eq.~(\ref{eqn:tineq}) then turns out to be equivalent to
\begin{equation}
 a \frac{1-\eta}{1+\eta} \; \left [ 1 - \eta^{2\kappa} \right ] < 2a - \mu 
<  a \frac{1-\eta}{1+\eta} \; \left [ 1 - \eta^{2(\kappa+1)} \right ].
\label{eqn:modeintervals}
\end{equation}
The disjoint open intervals defined above, together with their closure points, 
cover the range $0 < 2a - \mu < a (1-\eta)/(1 + \eta)$, 
which is precisely the mode-locking region for the steady state flows with $s = r = 1$ \cite{Aubry1983a} for a given $\eta$. 
In Ref.~\cite{Aubry1983a} this interval was calculated as the range of values of $\mu$ for which the energy per particle in the lowest energy configuration with a given average spacing is minimum.  In the present framework however, this interval arises from a restriction 
on the form of the flow pattern at steady-state.

Immediately to the left (right) of the points $\mu_\kappa$ defined as 
\begin{equation}
2a - \mu_\kappa = a \frac{1-\eta}{1+\eta} \; \left [ 1 - \eta^{2\kappa} \right ],
\end{equation}
the steady-state profile contains one more (less) secondary shock, while at $\mu_\kappa$ the 
global shock and its adjacent secondary shock collide at the time of the shock insertion. 
Within each open interval of Eq.~(\ref{eqn:modeintervals}) the steady state flow pattern has $\kappa + 2$ 
shocks. We thus see from Eq.~(\ref{eqn:xiglobal}) that as $2a - \mu$ approaches the phase boundary 
$2a - \mu_\infty = a (1-\eta)/(1+\eta)$, there is an accumulation of infinitely many shocks at the global shock whose location $\tilde{\xi}^{(\infty)}_0 \rightarrow 0$. Notice that this is also the location of the 
particles in the corresponding lowest energy configuration, confirming  
the result that at the phase boundaries the trajectory of the global shock coincides with 
that of the global minimizer \cite{BecKhanin}.

Finally, let us obtain from the inequality Eq.~(\ref{eqn:modeintervals}) a bound on the location of the global shock $\tilde{\xi}^{(\infty)}_0$ . The result is 
\begin{equation}
a \eta^{\kappa + 1} < \tilde{\xi}^{(\infty)}_0 < a \eta^{\kappa}.
\end{equation}
Note that the boundaries of the above intervals are the possible locations of secondary shocks, 
Eq.~(\ref{eqn:xisec}).

\begin{figure*}[t]
\begin{center}
\includegraphics[scale=0.58]{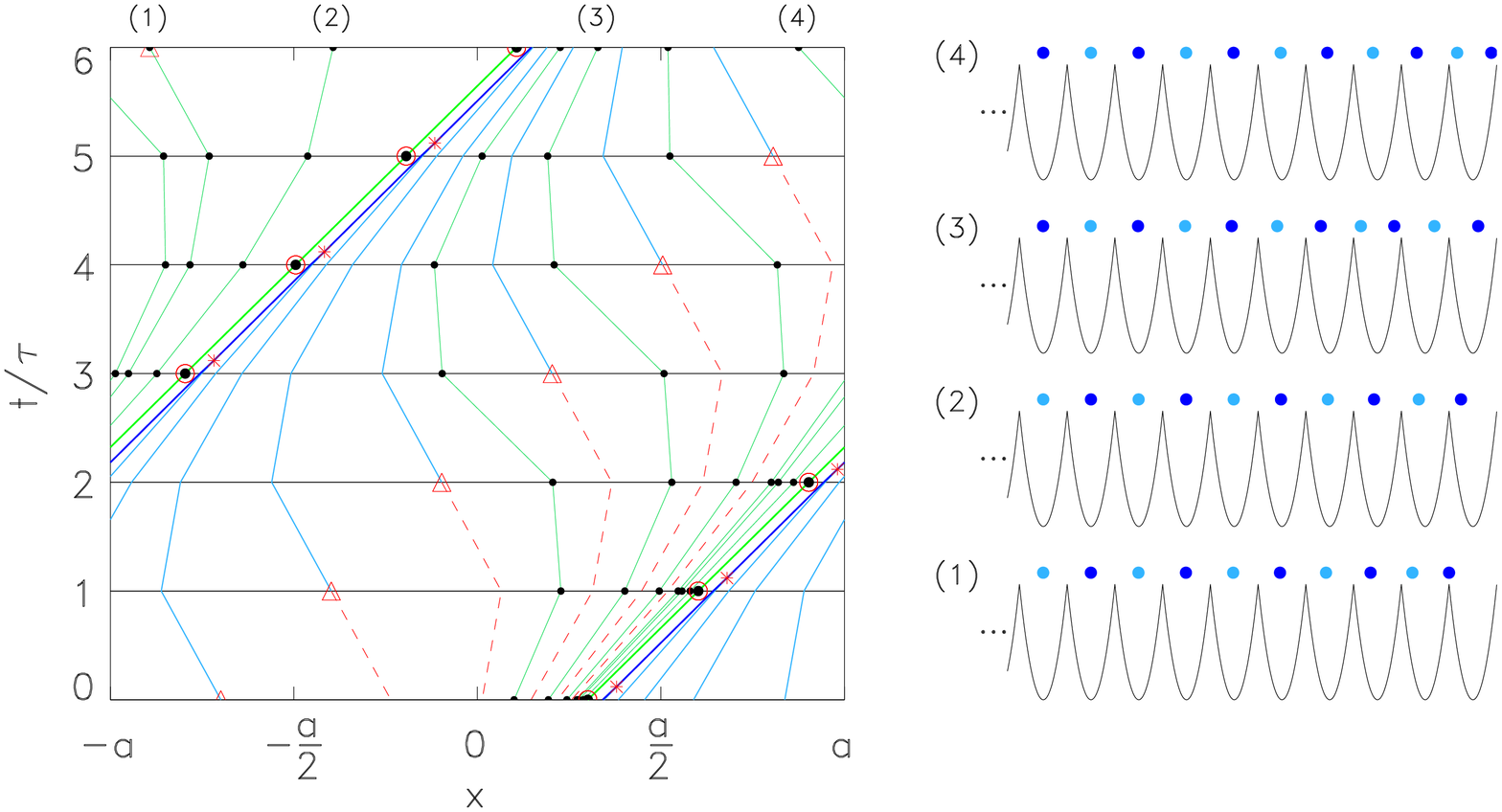}
\end{center}
\caption{(Color online) Sample particle configurations of a semi-infinite chain for  $\mu/2a = 0.8945$ and  $\lambda_0 = 0.4$, corresponding to $\ell/2a = 1/1$. Left: The light green line corresponds to the global minimizer, the lowest energy configuration. The darker green lines are a subset  of one-sided minimizers. These correspond to 
lowest energy configurations of the semi-infinite chain with the position of the outermost particle fixed. 
Particle locations are shown as black circles. Global and secondary shocks are shown in dark and light blue, respectively. Right: The particle configurations relative to the 
external potential for the corresponding one sided minimizers on the left panel labeled as (1) - (4). 
Notice the presence of discommensurations in (3) and (4), where the unit cell contains an additional 
particle. The alternate coloring of the particles shows that when a discommensuration occurs, all the particles to the left of the discommensuration must have moved by a period $2a$. The discommensurations occur precisely when the backwards flow of the characteristics changes 
from the right to the left half of the unit cell. The characteristics shown in red terminate at a 
newly inserted shock, they are known as {\it pre-shocks}.}
\label{fig:dislocation_example}
\end{figure*}

\subsection{One-Sided Minimizers and Discommensurations}
\label{disco}

The flow pattern also reveals what happens if we pick a particular time $t$ and look at the configurations 
of the semi infinite chain as the endpoint $x$ moves from $-a$ to $a$. The locations of the 
particles with respect to the unit cell can be read-off by noting where the corresponding location on the one-sided minimizer lies with respect to the well minima (red circles) and the well boundaries (red triangles). 
Sample one-sided minimizers and the corresponding particle configurations are shown in Fig.~\ref{fig:dislocation_example} for the one-periodic case $s=r=1$.

As we move into the chain, $t \rightarrow -\infty$, the coordinate of the
corresponding particle in the external frame has to decrease. For the one-sided minimizers, 
{\it i.e.} the flow of characteristics traced backwards in time, this means
that whenever the particle location relative to the unit cell increases, it
must necessarily have changed wells. However, a decrease in the coordinate, 
in particular a move from the right half
$(0,a)$ to the left half $(-a,0)$ of the unit cell, implies that the particle
remains in the same well and creates a {\it discommensuration} in the case
$s=r=1$.  

The case of flow patterns with multiple shock trees is similar. 
In general, by keeping track of the number of insertions after which a change of unit cell occurs, one can determine if the given potential well contains additional or missing particles, {\it i.e.} discommensurations. 
One-sided minimizers flowing in the (global) regions between the shock trees asymptotically approach the lowest energy particle configurations without 
incurring any discommensurations. On the other hand, one-sided minimizers starting in a region between the branches of a shock tree, upon entering the global region in between the trees, generate discommensurations. The boundary separating these two type of regions is a {\it pre-shock} \cite{BecKhanin}, a characteristic 
that evolves into a shock, such as the dashed red lines in Fig.~\ref{fig:dislocation_example}.

The life-time of a newly inserted shock, namely the number of insertions before it merges with the global  shock, also corresponds to the maximum number of particles counting from the end point of the 
chain within which a discommensuration can occur. As we have shown for the period-one case, the phase boundary of the corresponding domain in the $\mu-\eta$ plane is marked by an infinite number of secondary shocks. 
This turns out to be true for the phase boundaries of all the domains $\ell/2a = r/s$\cite{BecKhanin} and  implies that there are particle configurations with discommensurations arbitrarily deep inside the chain. 

We now turn to the calculation of the particle configurations associated with the minimizers for 
the case $s=r=1$. Denote by 
$I_k$ the regions bounded by the secondary shocks, $I_k = (\tilde{\xi}_0^{(k)}, \tilde{\xi}_0^{(k-1)})$ for 
$k = 1, 2, \ldots, \kappa $, which using Eq.~(\ref{eqn:xisec}) is given by
\begin{equation}
 I_k = (a \eta^{k}, a \eta^{k-1}).
\end{equation}
Likewise denote by $I_{\kappa+1} = (\tilde{\xi}_0^{(\infty)}, \tilde{\xi}_0^{(\kappa)})$ the region bounded by the global shock and the left-most secondary shock. The one-$\tau$ backwards flow maps $I_k$ into 
$I_{k-1}$. Let the initial time be $t_0 = 0$. The interval $I_k$ is terminated 
on its right end by $\tilde{\xi}^{(k-1)}$ so that the segment of $u$ is $u^{(k-1)}_{-}(\tilde{y}) \equiv u^{(k-1)}(\tilde{y}, 0^-)$, given as 
\begin{equation}
 u^{(k-1)}_{-}(\tilde{y}) = \lambda^*_+ \left ( \tilde{y} - \tilde{\nu}^{(k-1)}_0 \right ) - \lambda_0 \tilde{y}, 
\end{equation}
which noting that $\lambda^*_-\tau = \lambda^*_+ \tau - \lambda_0 \tau$,  $\lambda^*_-\tau = 1 - \eta$, 
and $\lambda^*_+ \tau = (1-\eta)/\eta$ can be rewritten as 
\begin{equation}
 u^{(k-1)}_{-}(\tilde{y}) \tau = (1 - \eta) \left ( \tilde{y} - \frac{\tilde{\nu}^{(k-1)}_0}{\eta} \right ).
\end{equation}
The $\Delta t = \tau$ backwards flow of the characteristics maps $\tilde{y}_{k} \in I_k$ into $x_{k-1}$ as
\begin{equation}
 x_{k-1} = \tilde{y}_{k} - u^{(k-1)}_{-}(\tilde{y}_{k} )\tau.
\end{equation}
Expressing $x_{k-1}$ in the unit cell coordinates, as $\tilde{y}_{k-1} = x_{k-1} + 2a - \mu$ and using Eq.~(\ref{eqn:nutilde_k}) we obtain the backwards recursion for the minimizer
\begin{equation}
 \tilde{y}_{k-1} = \eta \tilde{y}_{k} + 2a ( 1-\eta)\eta^{k-1},
\end{equation}
with $\tilde{y}_{k} \in I_k$. The solution is found as 
\begin{equation}
\tilde{y}_{k-j} = \eta^j \tilde{y}_{k} + \frac{2a}{1 + \eta} \; \eta^k \; \left ( \eta^{-j} - \eta^j \right ) - 2a \delta_{jk}.
\label{eqn:onesided1}
\end{equation}
This equation is valid for $j = 0, 1, 2, \ldots, k$. The case $j = k$ corresponds to the transition 
from the right to the left half of the unit cell. It can be shown that in order to bring the 
coordinate back into the unit cell an additional $2a$ has to be subtracted, accounting for the last term 
in the above equation.

Denote by $I_0$ the interval between the left boundary of the unit cell and the global shock, 
\begin{equation}
I_0 = [-a,\tilde{\xi}^{(\infty)}_0). 
\label{eqn:I0}
\end{equation}
As is apparent from Figs.~\ref{fig:dislocation_example} and \ref{fig:eql1o1}, for $\tilde{y}_0 \in I_0$, 
it must be that $\tilde{y}_k \in I_0$ for all $k < 0$. We will now verify this explicitly. Notice 
that $I_0$ is the interval belonging to the global segment of $u$. The corresponding intercept 
in the unit cell coordinates is given by $\tilde{\nu}^{(\infty)}_0$, Eq.~(\ref{eqn:nutilde_k}). 
Working again in the unit cell coordinates, the backwards map for 
$k \le 0$ turns out to be 
\begin{equation}
 \tilde{y}_{k-1} = \eta \tilde{y}_k
\end{equation}
whose solution is given as 
\begin{equation}
\tilde{y}_{-k} = \eta^k \tilde{y}_{0}\;\;\;\;\;\; k \ge 0.
\label{eqn:onesided2}
\end{equation}

It is clear that as $k \rightarrow \infty$, $\tilde{y}_{-k} \rightarrow 0$
monotonously, thus converging 
to the lowest energy configuration. Combining Eqs.~ (\ref{eqn:onesided1}) and (\ref{eqn:onesided2}), we have thus explicitly shown that all one-sided minimizers converge to the global minimizer $\tilde{y} = 0$.

The coordinates $y_j$, $j \le k$ of the corresponding configuration in the fixed frame turn out to be given as 
\begin{equation}
y_j = \left \{ \begin{array}{ll}  \tilde{y}_j + (j-1) 2a, & 2 \le j \le k \\ 
\tilde{y}_j, & j = 0, 1 \\
\tilde{y}_j + 2a j, & j < 0
\end{array} \right.
\label{eqn:disco2}
\end{equation}
with $\tilde{y}_k \in I_k$. 
The discommensuration is generated by the particles $j = 0$ and $1$, which are in the same cell. 

For a bi-infinite chain with particle $k$ fixed at $y_k$ such that the 
corresponding unit cell coordinate satisfies $\tilde{y}_k \in I_k$ with $k > 0$, the corresponding 
configuration still contains only a single discommensuration. Note that the 
other semi-infinite half extending to the right is equivalent to a semi-infinite chain extending to the 
left with its end-point at $-\tilde{y}_k$ and the chain reflected around the axis $\tilde{y} = 0$. 
Due to the structure of the intervals $I_k$, it follows that if $\tilde{y}_k \in I_k$ with $k > 0$ 
then $-\tilde{y}_k \in I_0$. Hence if $\tilde{y}_k \in I_k$ with $k > 0$, the semi infinite chain extending to the right 
cannot contain any additional discommensuration. The bi-infinite chain contains thus a 
single discommensuration. 

There are also bi-infinite chain configurations 
without any discommensurations. They are given by 
$\tilde{y}_0 \in (-\tilde{\xi}^{(\infty)}_0, \tilde{\xi}^{(\infty)}_0)$. This is the region 
between the global shock and its image obtained upon reflection at $\tilde{y} = 0$. Note in particular 
that the lowest energy configuration generated from $\tilde{y}_0 = 0$ belongs to this interval as well, 
as it should.

\section{Discussion}

We have shown that the flow of characteristics associated with a forced inviscid Burgers equation  
is related to the lowest energy configurations of FK chains. The trajectories of these characteristics traced 
backwards in time are the minimizers: the one-sided minimizers generate the lowest energy 
configuration of a semi-infinite chain for which the location of the outermost particle is fixed. 
They also converge to limiting trajectories, the global minimizers, which generate the 
lowest energy configurations of the bi-infinite chain. The flow 
of minimizers is confined to channels bounded by the trajectories of shock discontinuities that emerge from Burgers evolution. The shocks form tree-like structures and separate topologically distinct configurations of the 
semi-infinite chain that are marked by the presence or absence of discommensurations and their locations.
The shocks and their evolution are a consequence of the weak solutions to Burgers equation and, as we 
have shown, follow from thermodynamical considerations.

There are possible extensions of the approach presented here. Flow patterns containing shocks 
imply that the corresponding particle configurations are pinned by the external potential. 
In fact, the case of a piece-wise parabolic potential can be considered as the limit when the external 
potential is so strong that particles are mostly confined to the bottom of the 
potential wells, which can be approximated by parabolic segments. It is therefore natural to consider  
potentials that deviate from being piece-wise parabolic. 
As is clear from our results, the corresponding profiles will still consist of continuous segments  terminated by shock discontinuities, but the segments will not be 
straight lines anymore and the flow pattern will be perturbed. It should be possible to carry out a perturbation calculation. Knowing where this might break down would shed 
further light on the relation between shapes of external potentials and the intricate phase diagrams for 
the structure of their lowest energy configurations. 
   
For steady-state flow 
patterns containing shocks, the corresponding Burgers profile will always contain a segment that is 
bounded by shocks that will never merge and thus the segment will never disappear. In fact, the 
existence of such a global segment is guaranteed under rather general conditions \cite{EKMS,E}. 
Since the global segment contains the global minimizers, we were able to obtain these by simply searching 
for the characteristic trajectories in this region having the appropriate periodicity. 
The flow in the global region also allows us to calculate 
the backwards flow of characteristics in the vicinity of the global minimizer to which they necessarily converge. However, as long as the location of the shocks marking the boundary of the global region are not known, it cannot be asserted whether these characteristics are genuine one-sided minimizers and thus correspond to lowest energy configurations of the semi-infinite chain or not. To give an example, 
without the knowledge of the locations of shocks in Fig.~\ref{fig:dislocation_example}, the 
corresponding particle configurations on the right panel cannot be determined. 

Thus while it appears to be possible to calculate the global minimizers from limited local information 
of the flow, in order to calculate one-sided minimizers we require the full flow pattern including shocks. 
This corresponds to the limited extent of Aubry's theorem prescribing only the structure of the lowest energy configurations associated with the global minimizers, but not those associated with the one-sided minimizers. For systems with random external potentials, for which Aubry's theorem is not  
applicable and an exact analytical treatment might not be possible, 
one could still be able to determine the global minimizers even if only 
approximately.
Such an approach is implicit in the work of Feigel'man~\cite{Feigelman}, 
where a description similar to a periodically-forced Burgers equation was constructed for a charge-density wave system with random impurities with a focus on calculating the effective impurity pinning strengths rather than the phase configurations.  

As we have shown, the description in terms of a periodically forced Burgers equation lends itself to including temperature, Eq.~(\ref{eqn:viscidBurgers}). The evolution Eq.~(\ref{eqn:ForcedBurgers}) becomes now  
\begin{equation}
 u_t + u u_x = \frac{kT}{2} \; u_{xx} \; + \; \sum_{n = 0}^{\infty} \delta(t - n\tau) \; V^\prime(x+n\tau),
\nonumber
\end{equation}
where $u(x,t)$ is related to the free-energy $\epsilon(x,t)$ of the semi-infinite chain as $\epsilon_x(x,t) =  u(x,t)$. 
Note that for non-zero temperatures 
the viscous term smoothens out $u(x,t)$. 
In the case of a piece-wise parabolic potential for which 
$u(x,t)$ consists of linear segments, the primary effect of $T$ will be a rounding of the discontinuities marking 
its boundaries, while the interior of the segments will still remain approximately linear. 
Dimensional analysis shows that the length scale over which the shock discontinuity is smoothened out is 
of the order $\Delta_T \sim \sqrt{kT\tau}$. Thus one expects that regions with shock  spacing of 
order $\Delta_T$ or less will coalesce. This can happen at the accumulation of shocks in the profile of $u(x,t)$ near the phase boundary of a domain with a given $\ell$, as well as when the trajectory of a minimizer flows close to a shock, such as the global minimizer in Fig.~\ref{fig:dislocation_example}. As we have 
seen, for both of these cases the distances are of the order $\delta \sim a\eta^\kappa$. Thus 
when $\Delta_T$ and $\delta$ are comparable we expect that the corresponding configurations  will be susceptible to thermal fluctuations, that can give rise to discommensurations. On the other hand, for 
those portions of the minimizer that stay sufficiently far from shocks $(\delta \gg \Delta_T)$ the segments of $u$ remain still approximately linear, and they should therefore be less prone to thermal fluctuations. The discommensurations formed under  
thermal fluctuation were prescribed in \cite{VAS} and are precisely of the form given in Eq.~(\ref{eqn:disco2}). This is what one would expect, if the temperature is sufficiently small 
so that the density of discommensurations is low. 

{\it Acknowledgments---} MM would like to acknowledge useful suggestions at the early  
stages of the work from Susan~N.~Coppersmith and Valerii~M.~Vinokur, 
as well as later discussions with Paul~B.~Wiegmann, Konstantin Khanin, Serge Aubry and 
M.~Carmen Miguel. This research has been partly funded by Bo\u gazi\c ci University 
Research Grant 08B302. 

\appendix

\section{Characteristic Flows and the Inviscid Burgers Equation}
\label{onBurgers}

In this appendix we briefly review the weak solutions 
of the inviscid Burgers equation. For a more 
detailed account see \cite{Leveque, Whitham, Evans}.

Eq.~(\ref{eqn:burgers}) is in the form of a hyperbolic conservation law 
\begin{equation}
u_t + \left ( \frac{1}{2} u^2 \right)_x = 0.
\end{equation}
We are looking for a solution of  
\begin{equation}
u_t + u u_x = 0
\label{eqn:Fgen}
\end{equation}
subject to the initial condition 
\begin{equation}
u(x,0) = u_0(x).
\end{equation}

Given Eq.~(\ref{eqn:Fgen}), we define its characteristics as the curves $x(t)$ in the 
$xt$ plane on which $u(x,t)$ remains constant. These 
curves are straight lines given by the {\em characteristic equation}
\begin{equation}
x(t) = x_0 + t u_0(x_0),
\label{eqn:characteristics}
\end{equation}
with $u_0(x_0)$ being the speed of the characteristic emerging from the point $x_0$.
An implicit solution is then found as 
\begin{equation}
u(x,t) = u_0(x_0),
\label{eqn:Fconst}
\end{equation}
where  for a given $(x,t)$, $x_0$ is determined from the characteristic equation,  Eq.~(\ref{eqn:characteristics}). 

Depending on the initial conditions, the characteristics can  intersect, giving rise to multiple-valued points that are resolved by 
introducing discontinuities (shocks). Even with smooth initial data, discontinuities can 
develop in a finite time. Since solutions with discontinuities 
do not form a strict solution of the partial differential equation, one denotes these
as weak solutions which, instead of the local PDE, are required to obey 
a weaker form of the conservation law,  
\begin{equation}
\int_{0}^{\infty} {\rm d}t \int_{-\infty}^{\infty} {\rm d}x \; \chi(x,t) \left [  \frac{\partial u}{\partial t} + u  \frac{\partial u}{\partial x} \right ]   = 0,
\label{eqn:weak}
\end{equation}
for any continuously differentiable function $\chi(x,t)$ with compact support \cite{Leveque, Whitham, Evans}. This still does not uniquely determine the behavior of discontinuities. 
In general, this requires inspecting the 
microscopic evolution from which the continuum description arose. In the case of 
the mass-spring system of Section \ref{ssec:Burgers}, the weak solutions follow from demanding that the internal energy $H_{\rm int}(x,t)$ as a function of the end point of the spring is continuous.

Given a discontinuous segment of $u$ with the discontinuity at $x_0$, the
speed of the characteristics immediately to the left and right of $x_0$ are given as $u_l = u(x_0^{-},t)$ and $u_r = u(x_0^{+},t) $, respectively. 
There are two cases that one needs to distinguish:
(i) $u_l > u_r$, and (ii) $u_l < u_r$. 
In the former case, we have a moving shock discontinuity, while in the latter 
case, we have a {\it rarefaction wave}.  

{\bf (i) $u_l > u_r$:} Applying the integral form of the 
conservation law around the discontinuity, it can be shown that the shock 
moves with a speed 
\begin{equation}
 v = \frac{1}{2} \left (u_l + u_r \right ). 
\label{eqn:RH_condition}
\end{equation}
This is known as the Rankine-Hugoniot jump condition. 

{\bf (ii) $u_l < u_r$:} In this case the characteristics immediately 
to the left and right of the discontinuity at $x_0$ diverge from each other. The weak solution
in this case turns out to be given by  
\begin{equation}
u(x,t) = u_l + \frac{x-x_0}{t},
\label{eqn:simsol}
\end{equation}
for $(x,t)$ such that
\begin{equation}
0  < \frac{x - x_0}{t} <  u_r - u_l,
\end{equation}

\subsection{Shock Motion and Collisions}
\label{ShockDetails}

For the FK model with piece-wise parabolic potential, $u(x,t)$ is a series of 
straight line segments of identical slopes and discontinuities, as shown in 
Fig.~\ref{fig:representation}.
Consider first the evolution of a single straight line with initial slope $\lambda_0 > 0$ and 
$x$ intercept $\nu$, so that 
\begin{equation}
 u_0(x) = \lambda_0 (x - \nu).
\end{equation}
Characteristic lines emerging from $x_0$ move towards the left (right) for $x_0 < \nu$ ($x_0 > \nu$)
while the characteristic line emerging from $x_0 = \nu$ remains stationary. From the characteristic 
equation we thus find that 
\begin{equation}
u(x,t) = \lambda (t) \left ( x - \nu \right )
\label{eqn:uline} 
\end{equation}
with 
\begin{equation}
\lambda (t) = \frac{\lambda_0}{1 + \lambda_0 t}. 
\label{eqn:lambdat}
\end{equation}

If instead of a straight line we 
consider a line segment initially bounded by $x_l < x_r$: the 
evolution of this segment will again be given by Eq.~(\ref{eqn:uline}) with the restriction $x_l + u_0(x_l)t \le x \le  x_r + u_0(x_r)t$ and regardless of whether  $\nu$ lies 
inside or outside the interval bounded by $x_l$ and $x_r$.

We consider next the motion of a single shock  initially at $\xi_0$ such 
that the slopes of the segments immediately 
to its left and right are given by $\lambda_0 > 0$. Let $\nu$ and $\nu^\prime$ denote 
the locations where the segments to the left and right of the discontinuity intersect 
the $x$-axis. In order to have a shock discontinuity we also require that $\nu \le \nu^\prime$ and 
$u_0(x)$ is given as 
\begin{equation}
u_0(x) = \left \{  \begin{array}{cc} \lambda_0 (x - \nu) & x < \xi_0
  \\ \lambda_0 (x - \nu^\prime) & x > \xi_0 \end{array}   \right. . 
\end{equation}
The values of $u$, immediately to the left and right of the shock are 
\begin{equation}
u_l = \lambda_0 (\xi_0 - \nu) \ge u_r = \lambda_0 (\xi_0 - \nu^\prime) 
\end{equation}
and the initial speed of the shock is thus given by Eq.~(\ref{eqn:RH_condition})
\begin{equation}
v_0 = \lambda_0 \left ( \xi_0 - \frac{\nu + \nu^\prime}{2}  \right ).
\label{eqn:sxi}
\end{equation}

\begin{figure}[t]
\begin{tabular}{l} 
\includegraphics[height=5.7cm]{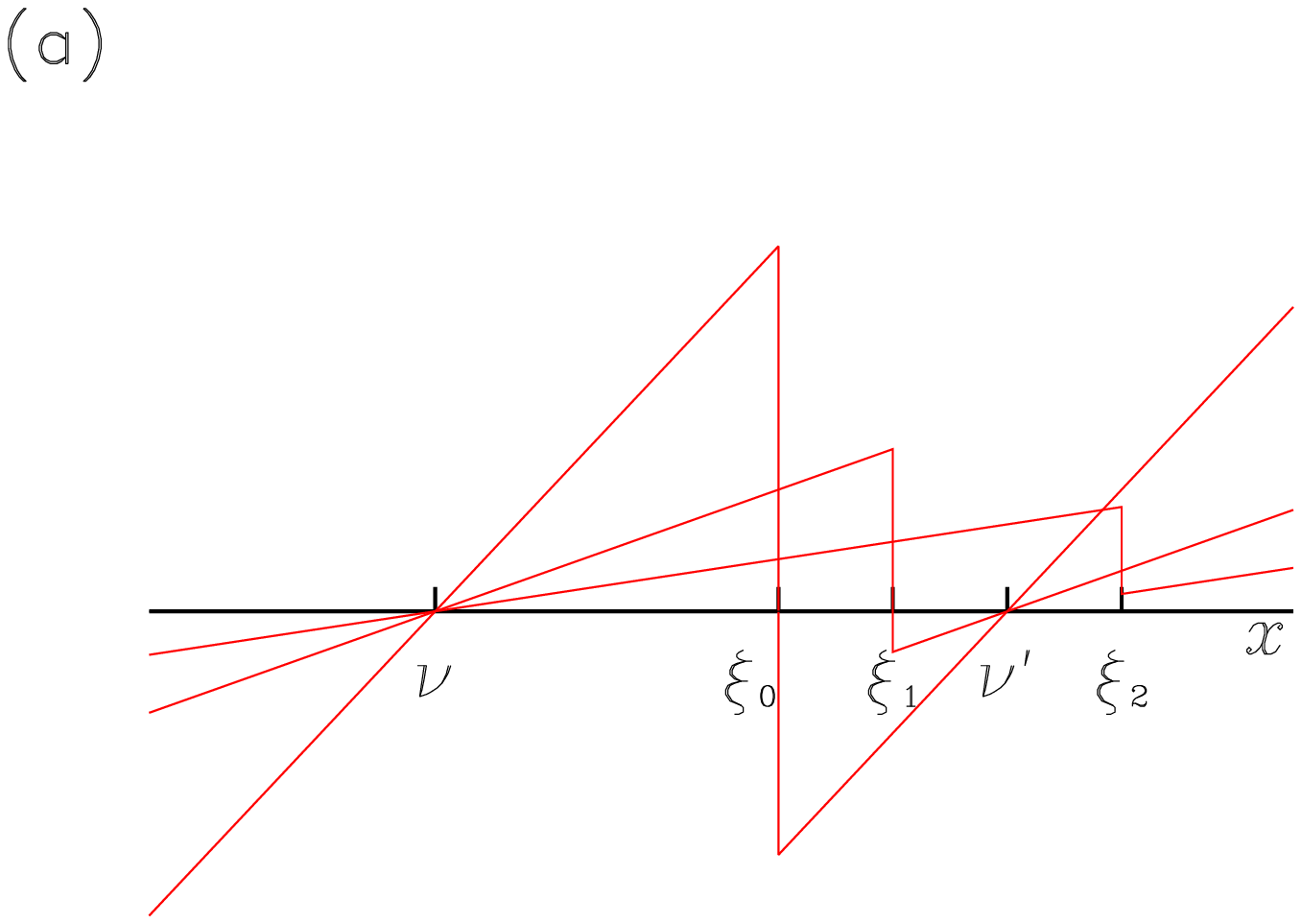}
\\
\includegraphics[height=5.7cm]{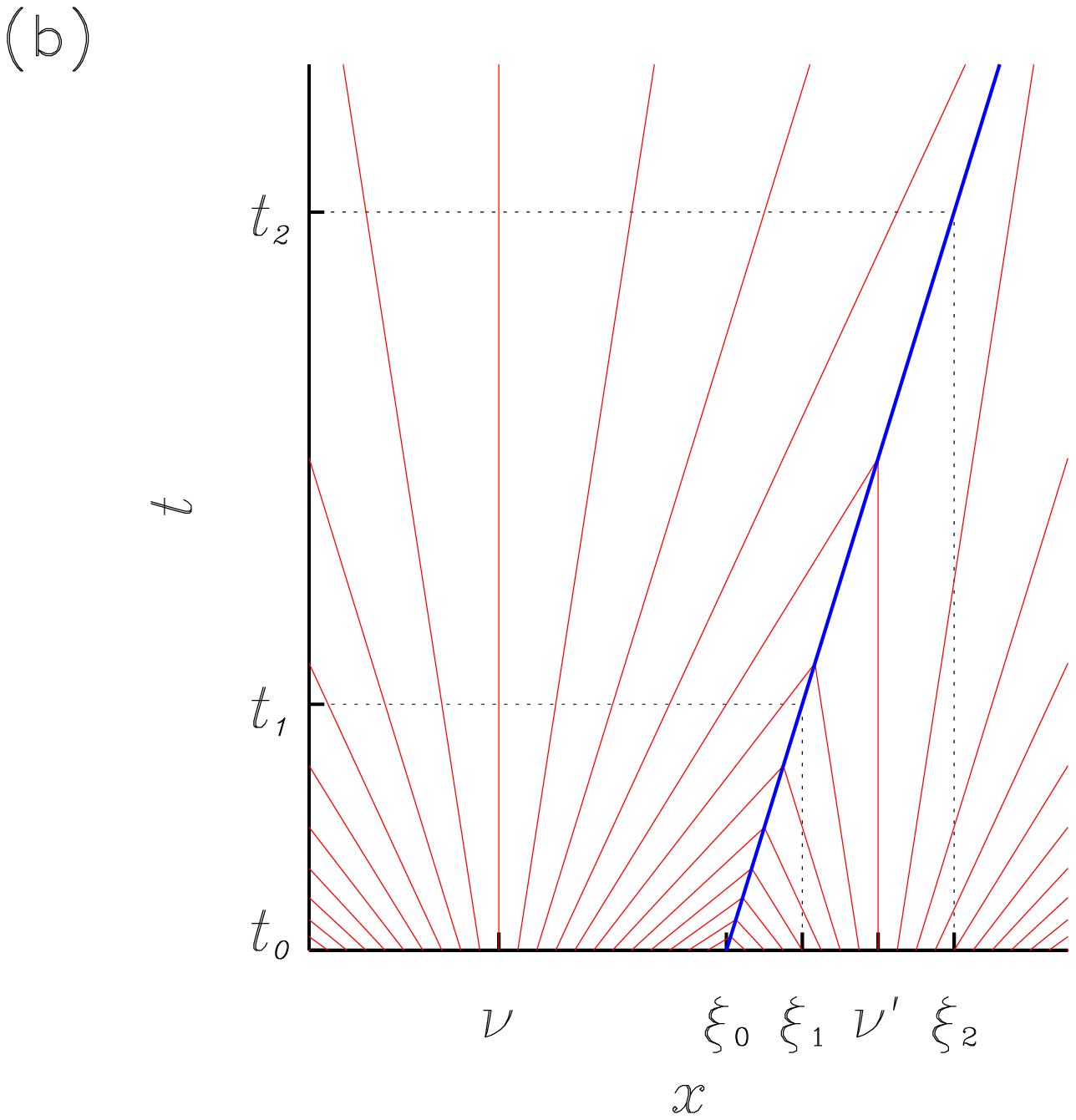}
\end{tabular}
\caption{(Color online) (a) Time evolution of the profile $u(x,t)$ containing a single shock. The intercepts of the left and right segments with the $x$ axis are denoted as $\nu$ and $\nu^\prime$. The position of the shock discontinuities  at subsequent times $t_0, t_1$ and $t_2$ are labeled by $\xi_0, \xi_1$ and $\xi_2$. (b) Characteristic lines associated with the profile (in red) and world-line trajectory of the 
shock discontinuity (in blue). Note how with increasing time more and more characteristics merge 
with the shock. }  
\label{fig:shock_example}
\end{figure}

From the definition of the shock speed Eq.~(\ref{eqn:RH_condition}), it is also clear that $u_l \ge v \ge u_r$. 
meaning that as time goes on, characteristic trajectories in the left and right vicinity of the 
shock will collide with the moving shock. For any given time $t$, the characteristics that have 
not yet collided with the shock will evolve their associated line segments according to 
Eq.~(\ref{eqn:uline}). As we have seen above, this evolution is such that the interception points 
$\nu$ and $\nu^\prime$ remain stationary. The slope of these segments will be given 
by Eq.~(\ref{eqn:lambdat}) and denoting by $\xi(t)$ and $v(t)$ the position and velocity of the 
shock at time $t$, respectively, we find that  
\begin{equation}
v(t) = \lambda (t) \left ( \xi (t) - \frac{\nu + \nu^\prime}{2} \right ).
\end{equation}
Differentiation of $v$ with respect to $t$ gives $\dot{v} = 0$, so that the shock moves at 
constant speed. Fig.~\ref{fig:shock_example} shows the evolution of $u(x,t)$ at three 
subsequent times $t_0,t_1$ and $t_2$ along with the flow of characteristics and the trajectory of the shock. 
In terms of the characteristic flow a shock acts like an attractor, gradually absorbing characteristics along with the associated values of $u_0(x)$ that flow with them.  
 
Consider now two shocks moving towards each other. We will denote the shocks 
as $\xi_l$ and $\xi_r$. The corresponding initial profile $u_0(x)$ will consist of three line segments, 
whose corresponding intercepts we will label as $\nu_l < \nu_m < \nu_r$. The initial profile is thus 
given as 
\begin{equation}
u_0(x) = \left \{ \begin{array}{ll}  
\lambda_0 (x - \nu_l) & x < \xi_l \\
\lambda_0 (x - \nu_m) &  \xi_l < x < \xi_r \\
\lambda_0 (x - \nu_r) & x > \xi_r 
\end{array} \right.
\end{equation}
while the corresponding speeds of the shocks are 
\begin{eqnarray}
v_l &=& \lambda_0 \left (\xi_l - \frac{\nu_l + \nu_m}{2} \right ) \;\;\;\;\; \mbox{and}  \;\;\;\;\; \\
v_r &=& \lambda_0 \left ( \xi_r - \frac{\nu_m + \nu_r}{2} \right ).
\end{eqnarray}
In order for the shocks to collide we must have that 
\begin{equation}
 \xi_l - \xi_r - \frac{\nu_l - \nu_r}{2} > 0.
\end{equation}

Let $t_c$ denote the time of collision.  As $t$ approaches $t_c$ the segment between the two 
shocks narrows until it disappears at $t_c$, leaving a single shock. The weak solution prescribes 
that for $t > t_c$ this shock will continue to move as a single shock with shock velocity $v^\prime$. 
By applying the Rankine-Hugoniot condition Eq.~(\ref{eqn:RH_condition}) at the instant the two shocks have 
just merged into a single shock, one finds that $v^\prime$ is given by
\begin{equation}
(\nu_m - \nu_l) v_l + (\nu_r - \nu_m) v_r = (\nu_r - \nu_l) v^\prime.
\end{equation}
The above equation resembles the conservation of momentum, and thus 
the two shocks behave like particles with (constant) ``masses'' $(\nu_m - \nu_l)$ and 
$(\nu_r - \nu_m)$ that collide inelastically. 

\section{Burgers Evolution of the FK Model with Piece-wise Parabolic Potentials}
\label{app:Details}

The variables $\nu^{(k)}$ and $\xi^{(k)}$ along with the asymptotic value of the profile slope $\lambda_+^*$ 
completely determine $u(x,t)$. During the evolution from $j\tau < t < (j+1)\tau$, the variables 
$\nu^{(k)}_j$ associated with segments $k$ remain constant or disappear due to merger of shocks, while the non-colliding shocks evolve according to Eq.~(\ref{eqn:xievol}). During the particle addition step
the shock locations $\xi^{(k)}_j$ remain unchanged as   
a new shock is inserted at $\xi^{\rm new}$, but the $\nu^{(k)}$ variables are mapped according to 
Eqs.~(\ref{nuupd}), (\ref{eqn:bi}) and (\ref{eqn:nunew}), while the parameter $b$ indicating the 
location of the zero intercept evolves according to Eq.~(\ref{eqn:b}). 
Together with the rules of how to handle colliding shocks, we thus have a discrete dynamical system 
for the variables $\nu$ and $\xi$ that underlies the evolution of $u(x,t)$ under the 
forced Burgers equation. Denoting by subscripts $j$ the times $t = (j\tau)^+$
right after shock insertion, the evolution equations for the segments that do
not disappear during a shock collision become

\begin{eqnarray}
b_{j+1} &=& b_j - \mu,  \label{eqn:eomfirst}\\
\nu^{(k)}_{j+1} &=& \eta \nu^{(k)}_j + (1 - \eta) b^{(k)}_j, \label{eqn:eomsecond}\\
b^{(k)}_j &=& \left \{ \begin{array}{ll} 
b_j, & \xi^{(k)} < \xi^{\rm new}_j \\
b_j + 2a, & \xi^{(k)} > \xi^{\rm new}_j \end{array} \right. , \label{eqn:eomthird} \\
\xi^{(k)}_{j+1} &=& \frac{1}{\eta} \xi^{(k)}_j - \frac{1- \eta}{\eta} \; \frac{\nu^{(k)}_j + \nu^{(k+1)}_j }{2}, \label{eqn:eomforth}\\
\xi^{\rm new}_{j+1} &=& \xi^{\rm new}_j - \mu. \label{eqn:eomlast}
\end{eqnarray}

The above equations assume that the segments $k$ are not involved in the collision of shocks. 
A segment $k$ will disappear during the time interval $[j\tau, (j+1)\tau)$ , if 
\begin{equation}
\Delta v^{(k)}_j  > 0 \;\;\;\;\mbox{and} \;\;\;\; \Delta \xi^{(k)}_j/ \Delta v^{(k)}_j < \tau, 
\end{equation}
where $ \Delta v^{(k)}_j \equiv v^{(k-1)}_j - v^{(k)}_j$ and $\Delta \xi^{(k)}_j \equiv \xi^{(k)}_j - \xi^{(k-1)}_j$. After the collision, $\xi^{(k)}$ will continue to move with a new velocity $s$ that 
has been worked out in \ref{ShockDetails}. 

\subsection{Feeding-Order of Newly Inserted Shocks and the Evolution of the Global Intercept}
\label{feeding-order}

The global intercepts lie in the strips bounded by shock trees and each of these strips 
contains one global minimizer. Thus at any 
time $t = n\tau$ there are $s$ locations of the global minimizers which correspond to the 
$s$ topologically distinct positions of the particles in the lowest energy 
configuration. 
Let us denote these locations by $\tilde{y}_\alpha$, with $\alpha = 0, 1, 2, \ldots , s-1$ and 
$\tilde{y}_\alpha \in [-a,a)$. Thus $\tilde{y}_\alpha$ are the locations of the particles in the 
external frame projected back into the unit-cell by translations of $2a$. 
The labeling is such that $\tilde{y}_0$ is the equilibrium configuration of the 
particle closest to the left boundary, $\tilde{y} = -a$, of the unit cell, $\tilde{y}_1$ refers to the particle in the lowest energy configuration immediately to its right, $\tilde{y}_2$ denotes its nearest next neighbor to the right {\em etc}. The labeling $\alpha$ is a numbering of the particles according to their positional order in the lowest energy configuration. Observe that unless $r=1$ the 
sequence $\tilde{y}_\alpha$ is not monotonously increasing, since the period of the 
configuration will comprise $r$ unit cells, whereas $\{ \tilde{y}_\alpha\} $ are the 
locations projected back into a single unit cell. 

Now focus on a single global minimizer. 
The location of this minimizer at a time $t = n\tau$ must correspond to one of the  
$\{ \tilde{y} \}$, say $\tilde{y}_\alpha$. Note that this location also  marks the position of 
the particle at the end point of a semi-infinite chain. At the next insertion time $t = (n+1)\tau$ the 
location of the global minimizer in the unit cell must necessarily be that of the next 
particle in the periodic configuration, say $\tilde{y}_\beta$. With the labeling convention given above, 
we have $\beta = (\alpha + 1) \;\;{\rm mod} \;\; s$. The same is true for all other global minimizers. Thus from  one insertion time to the next, the position of each of the global minimizers cycles through 
the ordered set $\{ \tilde{y}_\alpha \} $. 

On the other hand, at any given insertion time the locations 
of the $s$ global minimizers are distinct and they form the set $\{ \tilde{y}_\alpha\}$. Thus 
we can also order the set of $\{ \tilde{y}_\alpha\}$  according to proximity in the 
unit cell $[-a,a)$. Let us assume that the ordering in this way is given as   
$(\tilde{y}_{\alpha_0}, \tilde{y}_{\alpha_1}, \ldots, \tilde{y}_{\alpha_{s-1}})$, where 
$\alpha_0, \alpha_1, \ldots, \alpha_{s-1}$ is some permutation of $0,1, \ldots, s-1$.   
It is not difficult to convince oneself 
that the differences $(\alpha_i - \alpha_{i+1}) \;\; {\rm mod} \;\; s$ must be identical: 
Given a time $t=n\tau$, the location of 
the shock just inserted, $\xi^{\rm new}_n$, by definition also marks the left boundary of the 
unit cell. Thus $\tilde{y}_0$ defined above as the global minimizer closest to the left boundary 
is also closest to the new shock from the right.  
At time $t=(n+1)\tau$ a new shock is inserted at $\xi^{\rm new}_{n+1} = 
\xi^{\rm new}_n - \mu$, {\it cf.} Eq.~(\ref{eqn:eomlast}) and thus there is a corresponding 
global minimizer immediately to its right corresponding to $\tilde{y}_0$ at this new time.  
Thus when progressing in time, the location $\tilde{y}_0$ must cycle through the $s$ 
global minimizers which we had labeled as $\alpha = 0, 1, 2, \ldots, s-1$, at some earlier time 
$t_0 = n\tau$. The uniform shift by $-\mu$ of the location of the new shock to be inserted 
implies that this cycling of $\tilde{y}_0$ through the minimizers must also be a shift of the form 
$\alpha \rightarrow  (\alpha - \Delta) \;\; {\rm mod} \;\; s$, where $\Delta < s$ and $\Delta$ and $s$ are 
co-prime. In fact, $r \equiv \Delta$, so that this can be regarded as a definition of $r$. Thus 
for a steady state flow pattern corresponding to $\ell/2a = r/s$, $s$ determines the 
periodicity in time $s\tau$, while $r$ controls the ``feeding order'' of the shock trees. 

Observe now that the feeding order of the shock trees also determines whether the shock associated 
with the right boundary of a global segment is to the immediate left or right of the newly inserted shock 
in the co-moving coordinates: Recall that (i) 
to each inserted shock there corresponds a shock tree into which this shock will eventually flow, 
and (ii) that for any $t$, any two neighboring global minimizers are separated by a shock tree (and 
hence a global shock). The sequence of being to the left or right of the newly inserted shock must 
therefore also follow the feeding order. 

We thus find from  Eqs.~(\ref{eqn:eomfirst}) and (\ref{eqn:eomthird}) that for the global intercepts 
$\nu^{(k)}_j$ on a shock tree
\begin{equation}
b^{(k)}_j = -j\mu + 2a \; {\rm Int}\left [ (j+\delta)\frac{r}{s} \right ], 
\label{eqn:bkjelta}
\end{equation}
with each value of $\delta = 0, 1, 2, \ldots s-1$ being associated with one of the $s$ shock trees.


\begin{thebibliography}{10}
\bibitem{IC_Comm} In fact as we will see shortly, for sufficiently large times $t$, the flow reaches a steady-state limit in which the initial condition does not matter anymore.
\bibitem{oneshocktree} Only in the special case where at a given insertion time $t$ the shock tree contains only one shock, namely the global shock, the corresponding global intercept will also be the one associated with the global shock.
\bibitem{note} The case for $r=0$ and $s=1$ so that $\chi_0 = 0$ is analogous, but  
unphysical from a classical physics point of view, since the average spacing of particles in the lowest energy configuration vanishes. This results in a point-like condensate 
at the minimum of the potential well.
\end{thebibliography}
\end{document}